\documentclass[twocolumn,english,journal]{IEEEtran}
\usepackage[T1]{fontenc}
\usepackage{color}
\usepackage{babel}
\usepackage{calc}
\usepackage{units}
\usepackage{amsmath}
\usepackage{amssymb}
\usepackage{graphicx}
\usepackage[unicode=true,
 bookmarks=true,bookmarksnumbered=true,bookmarksopen=true,bookmarksopenlevel=1,
 breaklinks=false,pdfborder={0 0 0},backref=false,colorlinks=false]
 {hyperref}
\hypersetup{pdftitle={Your Title},
 pdfauthor={Your Name},
 pdfpagelayout=OneColumn, pdfnewwindow=true, pdfstartview=XYZ, plainpages=false}

\makeatletter
  \newtheorem{assumption}{Assumption}
  \newtheorem{definitn}{Definition}
  \newtheorem{lemma}{Lemma}
  \newtheorem{remrk}{Remark}
  \newtheorem{problem}{Problem}
  \newtheorem{thm}{Theorem}
  \newtheorem{example}{Example}

\usepackage[caption=false,font=footnotesize]{subfig}

\@ifundefined{showcaptionsetup}{}{%
 \PassOptionsToPackage{caption=false}{subfig}}
\usepackage{subfig}
\makeatother

\begin{document}

\title{MIMO Precoding for Networked Control Systems with Energy Harvesting
Sensors}

\author{Songfu Cai,\textsl{ Student Member, IEEE}, Vincent K. N. Lau, \textit{Fellow,
IEEE}\\
Department of Electronic and Computer Engineering \\
The Hong Kong University of Science and Technology\\
Clear Water Bay, Kowloon, Hong Kong \\
Email: \{scaiae, eeknlau\}@ust.hk}
\maketitle
\begin{abstract}
In this paper, we consider a MIMO networked control system with an
energy harvesting sensor, where an unstable MIMO dynamic system is
connected to a controller via a MIMO fading channel. We focus on the
energy harvesting and MIMO precoding design at the sensor so as to
stabilize the unstable MIMO dynamic plant subject to the energy availability
constraint at the sensor. Using the Lyapunov optimization approach,
we propose a closed-form dynamic energy harvesting and dynamic MIMO
precoding solution, which has an $\emph{event-driven control}$ structure.
Furthermore, the MIMO precoding solution is shown to have an $\emph{eigenvalue water-filling}$
structure, where the water level depends on the state estimation covariance,
energy queue and the channel state, and the sea bed level depends
on the state estimation covariance. The proposed scheme is also compared
with various baselines and we show that significant performance gains
can be achieved.\end{abstract}

\section{Introduction}

Networked control systems (NCSs) have become quite popular recently
due to their growing applications in industrial automation, smart
transportation, remote robotic control, etc.. A typical NCS consists
of a multiple-input multiple-output (MIMO) dynamic $\emph{plant}$
\textcolor{blue}{}\footnote{MIMO dynamic plant refers to the dynamic plant with vector state evolution
and vector control inputs \cite{ftnote1}.}, a multiple antenna wireless $\emph{sensor}$\textcolor{blue}{{} }\footnote{In some typical applications (such as 2.4GHz ISM WLAN and IEEE 802.15.4),
the number of available channels for NCS system operation can be very
limited \cite{ftnote1}. As such, multi-antenna sensor will be useful
to enhance the spectral efficiency despite the limited\textcolor{blue}{{}
}bandwidth in the system. }\textcolor{blue}{{} }and a $\emph{controller}$. These are connected
over a communication network and form a closed-loop control, as illustrated
in Figure 1. Compared to traditional wireless sensors that are powered
by a non-rechargeable battery, we consider an energy harvesting sensor,
which is equipped with an energy harvesting device (e.g., a solar
panel or a micro wind turbine) so that the sensor can harvest energy
from the surrounding environment. Such NCSs with energy harvesting
sensors have many advantages. For example, they do not require the
replacement of batteries (for traditional sensors) and the maintenance
process is simplified for NCSs in dangerous environments (e.g., chemical
plants).

For an NCS, a potentially unstable plant is stabilized using a sensor
to measure the state and feeds back to a controller, which in turn
drives a stabilizing control signal to the plant. Unlike conventional
feedback control design where the feedback path is assumed to be perfect,
the sensor in the NCS observes the plant state and transmits the observation
to a remote controller over a wireless MIMO fading channel. The performance
of the NCS is closely related to the communication resource control
at the energy harvesting sensor over the MIMO wireless channel. For
an energy harvesting sensor, the communication resource is related
to the energy available in the battery. Due to the random nature of
the renewable energy source, it is difficult to predict the future
evolution of the energy arrivals, and hence there is a tradeoff of
using the available energy to support the current transmission and
saving the energy for future good transmission opportunities. 

In fact, NCS is a multidisciplinary subject involving both control
theory and communication theory. The main objective of the control
theory is to control the plant such that its state evolves in a desired
manner. The plant is usually modeled as a linear dynamic system which
is described by a set of first order coupled linear difference equations
representing the evolution of the state variables. The representation
of a plant with linear system provides a convenient and compact way
to model and analyze the plant. Stability is an important characteristic
of control system. When a plant is unstable, the state of the plant
may be unbounded even though the input to the plant is bounded. General
uncertainties and external disturbances may cause the dynamic plant
to be unstable and this may incur costly physical damage. For instance,
an unstable aircraft may crash and an unstable chemical plant may
explode. Therefore, it is extremely important stabilize the unstable
dynamic plant. 

The theory of Lyapunov drift has a long history in the field of stochastic
control to analyze the system stability in the field of control and
communications. The authors of \cite{lya3} first applied the Lyapunov
drift theory to develop a general algorithm which stabilizes a multi-hop
packet radio network. The negative Lyapunov drift terms play a central
role when applying the Lyapunov drift theory to analyze the stability
of dynamic systems. Intuitively, the negative Lyapunov drift is a
stabilizing force that pulls the system state back to the equilibrium
point. In \cite{lya7}, it is shown that negative Lyapunov drift ensures
network stability because whenever the data queue length vector leaves
a certain bounded region, the negative drift eventually drives it
back to that region. In \cite{lya8}, the negative Lyapunov drift
is utilized to analyze the stability of Markov chains.

From the communication side, it is important to understand how to
optimize the communication resources (such as MIMO precoding) targeted
for control applications. In particular, MIMO precoder optimization
has been well studied in wireless communications. A MIMO precoder
is essentially a multimode beamformer which splits the transmit signal
into spatial eigenbeams and assigns higher power along the beams when
the channel is strong \cite{mmse-precoder}. In \cite{Sampath2001,wiesel2006linear},
the authors obtain the MIMO linear precoding solution to minimize
the MMSE \cite{Sampath2001} or maximize the SINR \cite{wiesel2006linear}
of the MIMO wireless systems. However, these precoding solutions are
not tailored for NCS applications because the optimization objectives
(MMSE or SINR) are merely physical layer metrics in wireless communications,
which may not be directly related to the performance metric of the
NCS applications (such as stability of the plant or plant state estimation
errors). Another related development in wireless communications is
the device-to-device (D2D) or machine-to-machine (M2M) communications
where a communication device can communicate and exchange information
with a peer device autonomously. One important application of D2D
or M2M systems (in addition to delivering content) is to support real-time
industrial control where the devices may act as sensors and/or actuators.
As such, these new application scenarios embrace both wireless communications
and control in the context of networked control systems. In this paper,
we are interested in studying the design of the MIMO precoder in the
communication subsystem to support NCS applications. 

Note that communication resource allocation for an NCS with an energy
harvesting sensor is quite challenging. This is because such a problem
embraces the information theory (to model the physical layer over
the wireless channels), the queuing theory (to model the energy queue
dynamics) and the control theory (to model the plant dynamics under
imperfect state feedback control). There are some works on the dynamic
resource control design for NCSs with an energy harvesting sensor.
In \cite{li2013optimal}, the authors study the mean square average
state estimation error minimization under renewable energy constraints.
To obtain the optimal communication power control policies, the associated
stochastic optimization problems are solved using the numerical value
iteration algorithm, which induces huge complexity, and suffers from
slow convergence and lack of insights \cite{bertsekas1995dynamic}.
In this paper, we propose a low-complexity closed-form dynamic MIMO
precoding solution at the sensor powered by renewable energy to stabilize
the unstable MIMO dynamic plant. The following summarizes the key
contributions in the paper. 
\begin{itemize}
\item \textbf{Direct Analog State Transmission with Peak Power Constraint:}
Unlike existing approaches in NCS \cite{event-driven1,event-driven2,optimal-kf-power-allo}
where the plant state is first quantized and transmitted over a simplified
digital channel, we propose a novel analog state transmission where
the sensor simply transmits a spatially rotated state measurement
(rotated by the MIMO precoder) to the remote controller without quantization
or coding. The consideration of renewable energy source also poses
a unique challenge. For instance, the sensor needs to equip with a
limiter, which is a non-linear module to attenuate peaks of signals
to satisfy the peak power constraint \footnote{The sensor cannot transmit more than the instantaneous available energy
in the battery.}. Such design will simplify the datapath design of the sensor and
the remote controller and directly take advantage of the MIMO communication
channels. 
\item \textbf{Closed-form Dynamic Event-Triggered MIMO Precoder Policy:}
Event-triggered sensor control has been proposed in various existing
NCS applications \cite{optimal-kf-power-allo,optimal-strategy}. However,
the existing solutions either has no closed-form triggering solution
\cite{optimal-kf-power-allo,optimal-strategy} or the solution is
not truly dynamic \cite{optimal-kf-power-allo}. In addition, in all
these existing solutions, there is no consideration of the multi-antenna
MIMO communication channels. In this paper, we derive a closed-form
fully dynamic solution which adapts to the complete system state (MIMO
channel state for transmission opportunity, sensor energy state for
energy availability and plant state estimation error for the urgency
of the state transmission). These agile adaptivity are very important
for superb performance in the NCS application. 
\item \textbf{Closed-form Performance Characterizations: }In this paper,
we also derive closed-form requirement of the renewable energy arrival
rate and the battery capacity requirement to attain stability of the
MIMO NCS. Such results give important guideline for the dimensioning
of the resource needed at the sensor. Furthermore, we have derived
closed form MSE and study how the performance depends on key system
parameters. Such closed-form derivation is challenging due to the
dynamic MIMO precoding policy as well as the coupled system state
evolutions where the dynamic evolution of the three system states
(the energy state, the plant state and the MIMO channel state) are
tightly coupled together in a very complicated manner. 
\end{itemize}

\textit{Notation}s: Uppercase and lowercase boldface denote matrices
and vectors, respectively. The operators $(\cdot)^{T}$, $(\cdot)^{\dagger}$,
$(\cdot)^{H}$, $\mathrm{Tr\left(\cdot\right)}$, $|\cdot|$, $\mathrm{Re}\left\{ \cdot\right\} $,
$\mathbf{1}{\{\cdot\}}$ are the transpose, element-wise conjugate,
conjugate transpose, trace, cardinality, real part, and indicator
function, respectively; $||\mathbf{A}||$ and $||\mathbf{a}||$ denote
spectrum norm of matrix $\mathbf{A}$ and Euclidean norm of vector
$\mathbf{a}$, respectively; and $\text{diag}(\textbf{a})$ means
diagonal matrix with diagonal elements being $\textbf{a}$; $\mathbf{A}_{ij}$
denotes the element in the $i$-th row and $j$-th column of matrix
$\mathbf{A}$; $\mu_{i}\left(\mathbf{A}\right)$ denotes the $i$-th
largest eigenvalue of matrix $\mathbf{A}$. $\mathbb{R}^{m\times n}$
($\mathbb{C}^{m\times n}$) represents the set of $m\times n$ dimensional
real (complex) matrices.

\section{System Model}

In this section, we introduce the model of the MIMO NCS with an energy
harvesting sensor, plant dynamics, MIMO channel model, energy queue
model, as well as the information structures at the sensor and the
controller.

\subsection{MIMO Networked Control System with an Energy Harvesting Sensor }

\begin{figure}[tbh]
\begin{centering}
\includegraphics[clip,width=1\columnwidth]{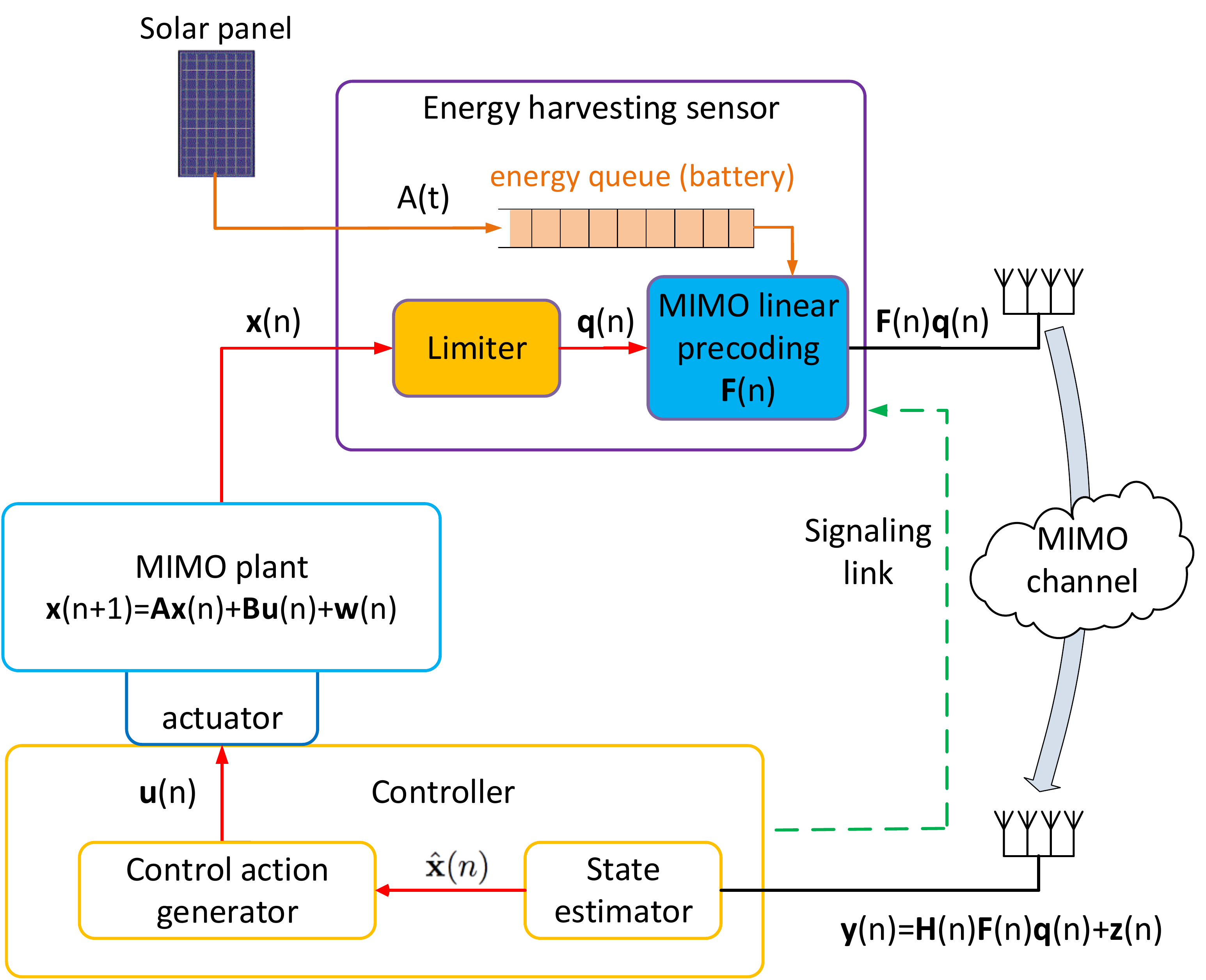}
\par\end{centering}

\caption{\label{fig:system set-up}A typical architecture of a MIMO NCS with
an energy harvesting sensor. In this paper, we focus on the dynamic
design of the limiter and MIMO precoder. }
\end{figure}

Figure \ref{fig:system set-up} shows a typical networked control
system (NCS) with a MIMO plant (potentially unstable), a multi-antenna
sensor (with $N_{s}$ transmit antennas) with energy harvesting capability
and a multi-antenna remote controller (with $N_{c}$ receive antennas).
The sensor and the controller are geographically separated and connected
through a wireless MIMO fading channel. We consider a time-slotted
system with slot duration $\tau$. The sensor has perfect observation
of the MIMO plant state $\mathbf{x}\left(n\right)$ at every time
slot. Due to the\textcolor{red}{{} }consideration of renewable energy
source at the sensor, the observed plant state $\mathbf{x}\left(n\right)$
is passed through an energy limiter before the transmission so as
to satisfy an instantaneous peak power constraint determined by the
instantaneous available energy. The output of the limiter is $\mathbf{q}\left(n\right)$.\textcolor{blue}{{}
}Unlike conventional digital approaches in NCS, we consider direct
analog transmission of the measured system state at the sensor. Specifically,
the analog state measurement is spatially rotated by a MIMO precoder
$\mathbf{F}(n)$ and transmits to the remote controller over the multi-antenna
wireless fading channel. We assume the perfect channel state information
can be obtained at the sensor by causal signaling feedback from the
controller as illustrated in Figure 1. The received signal at the
controller is $\mathbf{y}(n)$, which is passed to the state estimator
to obtain a state estimate $\hat{\mathbf{x}}\left(n\right)$. Based
on $\hat{\mathbf{x}}\left(n\right)$, the remote controller further
generates the control action $\mathbf{u}\left(n\right)$. The controller
is physically attached to the actuator of the plant so that the control
action $\mathbf{u}\left(n\right)$ generated at the controller can
be applied to the plant directly. The actuator, which is co-located
with the plant, then applies $\mathbf{u}\left(n\right)$ for plant
actuation. The goal of the MIMO NCS is to stabilize the potentially
unstable MIMO plant with limited wireless communication resources. 

\vspace{-0.2cm}

\subsection{Stochastic Dynamic MIMO Plant Model}

We consider a discrete-time stochastic MIMO plant system with state
dynamics: $\mathbf{x}\left(n+1\right)=\mathbf{A}\mathbf{x}\left(n\right)+\mathbf{B}\mathbf{u}\left(n\right)+\mathbf{w}\left(n\right)$,
$\ n\geq0$, $\mathbf{x}\left(0\right)=\mathbf{x}_{0}$, where $\mathbf{x}\left(n\right)\in\mathbb{R}^{K\times1}$
is the plant state process, $\mathbf{u}\left(n\right)\in\mathbb{R}^{D\times1}$
is the plant control action, $\mathbf{A}\in\mathbb{R}^{K\times K}$,
$\mathbf{B}\in\mathbb{R}^{K\times D}$, and $\mathbf{w}\left(n\right)\in\mathbb{R}^{K\times1}$
is the plant noise. We assume the plant noise $\mathbf{w}\left(n\right)\in\mathbb{R}^{K\times1}$
is zero mean with covariance $\mathbb{E}\left[\mathbf{w}\left(i\right)\left(\mathbf{w}\left(j\right)\right)^{T}\right]=\delta_{i,j}\mathbf{W}$
for $\forall i,j$, where $\delta_{i,j}=0$ if $i\neq j$ and $\delta_{i,j}=1$
otherwise. We assume that the plant noise covariance is finite \footnote{There exist a bounded constant $W$ such that $\mathbf{W}\leq W\mathbf{I}$.}.
The MIMO plant system $(\mathbf{A},\mathbf{B})$ is assumed to be
controllable with $\mathbf{A}$ containing possibly unstable eigenvalues\footnote{Unstable eigenvalues are eigenvalues with a modulus greater than 1
\cite{sastry2013nonlinear}.}.

\subsection{MIMO Wireless Channel Model}

We model the wireless communication channel between the multi-antenna
sensor and the controller as a wireless MIMO fading channel. Using
multiple-antenna techniques, the $N_{s}$-antenna sensor can deliver
$K$ parallel data streams to the $N_{c}$-antenna controller through
spatial multiplexing. We assume $K\leq\mathrm{min}\left\{ N_{s},N_{c}\right\} $.
At the $n$-th time slot, the received signal $\mathbf{y}\left(n\right)$
at the controller is given by 
\begin{equation}
\mathbf{y}\left(n\right)=\mathbf{H}\left(n\right)\mathbf{F}\left(n\right)\mathbf{q}\left(n\right)+\mathbf{z}\left(n\right),\label{asdaschhads}
\end{equation}
where $\mathbf{H}\left(n\right)\in\mathbb{C}^{N_{c}\times N_{s}}$
is the MIMO channel fading matrix, $\mathbf{F}\left(n\right)\in\mathbb{C}^{N_{s}\times K}$
is the MIMO precoding matrix, $\mathbf{q}(n)$ is the output of the
energy limiter\footnote{We shall illustrate the energy limiter structure in detail in Section
II-E and Section III-B.}, and $\mathbf{z}\left(n\right)\sim\mathcal{CN}\left(0,\mathbf{I}_{N_{s}}\right)$
is the additive complex Gaussian noise. We have the following assumption
on $\mathbf{H}\left(n\right):$

\begin{assumption}
\emph{(MIMO Wireless Channel Model)} The random MIMO channel realization
$\mathbf{H}\left(n\right)$ remains constant within each time slot.
Furthermore, $\mathbf{H}(n)$ is i.i.d. across different time slots
according to some general distribution.
\end{assumption}

\vspace{-0.2cm}

\subsection{Energy Harvesting and Energy Queue Model at the Sensor}

We assume the sensor is solely powered by renewable energy sources
(such as a solar panel). Due to the random nature of the energy source,
a battery or ultra-capacitor is needed to store the harvested energy
at the sensor. Let $\alpha\left(n\right)$ be the amount of harvestable
energy at the sensor at time slot $n$. We have the following assumption
on harvestable energy process $\alpha\left(n\right)$.
\begin{assumption}
\textsl{(Renewable Energy Model)} The harvestable energy $\alpha\left(n\right)$
is i.i.d. across different time slots according to some distribution.
\end{assumption}

The sensor has an energy storage (or battery) so let $E\left(n\right)$
be the amount of energy left in the storage device at time $n$. We
assume that the sensor is causal in the sense that new energy arrivals
are observed after the control actions are performed at each time
slot. Hence, the energy queue dynamics at the sensor is given by
\begin{align}
E\left(n+1\right) & =\mathrm{min}\left\{ \left[E\left(n\right)-\|\mathbf{F}\left(n\right)\mathbf{q}\left(n\right)\|^{2}\tau\right]^{+}+\alpha\left(n\right),\theta\right\} ,\label{eq:energy-dya}
\end{align}
where $\left[x\right]^{+}=\mathrm{max}\left\{ x,0\right\} $, $\|\mathbf{F}\left(n\right)\mathbf{q}\left(n\right)\|^{2}\tau$
measures the energy consumed at time slot $n$ for delivering $\mathbf{q}\left(n\right)$
using precoding action $\mathbf{F}\left(n\right)$ over the MIMO wireless
fading channel, and $\theta$ is the sensor battery capacity. Furthermore,
at any time slot $n$, the precoding control action must satisfy the
following $\emph{energy availability}$ constraint:
\begin{align}
\|\mathbf{F}\left(n\right)\mathbf{q}\left(n\right)\|^{2}\tau\leq E\left(n\right).\label{energyav}
\end{align}
\vspace{-0.8cm}

\begin{center}
\fbox{\begin{minipage}[t]{1\columnwidth}%
Challenge 1: Energy Availability Constraint and Saturation.%
\end{minipage}}
\par\end{center}

The $\emph{energy availability}$ constraint (\ref{energyav}) means
that the energy consumption at each time slot cannot exceed the current
available energy in the energy buffer at the sensor. Such a constraint
greatly complicates the design of the dynamic MIMO precoder at the
sensor. At any time slot, the dynamic MIMO precoder has to strike
a balance between how much energy to consume, the $\emph{good transmission opportunities}$
induced by the fading channel and the $urgency$ induced by the state
estimation errors at the controller. Furthermore, there is an effective
$\emph{peak transmission power constraint}$ at the sensor at each
time slot. In order to satisfy this constraint, the sensor needs to
be equipped with a $\emph{limiter}$. Since the input process $\mathbf{x}\left(n\right)$
is non-stationary with unbounded support, there is non-zero chance
that the input $\mathbf{x}\left(n\right)$ exceeds the limiter range,
causing $\emph{saturation}$. This causes non-linearity in the feedback
loop and substantially complicates the optimization problem. In the
next section, we introduce the design of the limiter, and we will
elaborate how we tackle the issue of saturation in Section III-B.

\vspace{-0.4cm}

\subsection{Energy Limiter at the Sensor}

\begin{figure}[tbh]
\begin{centering}
\includegraphics[clip,width=0.8\columnwidth]{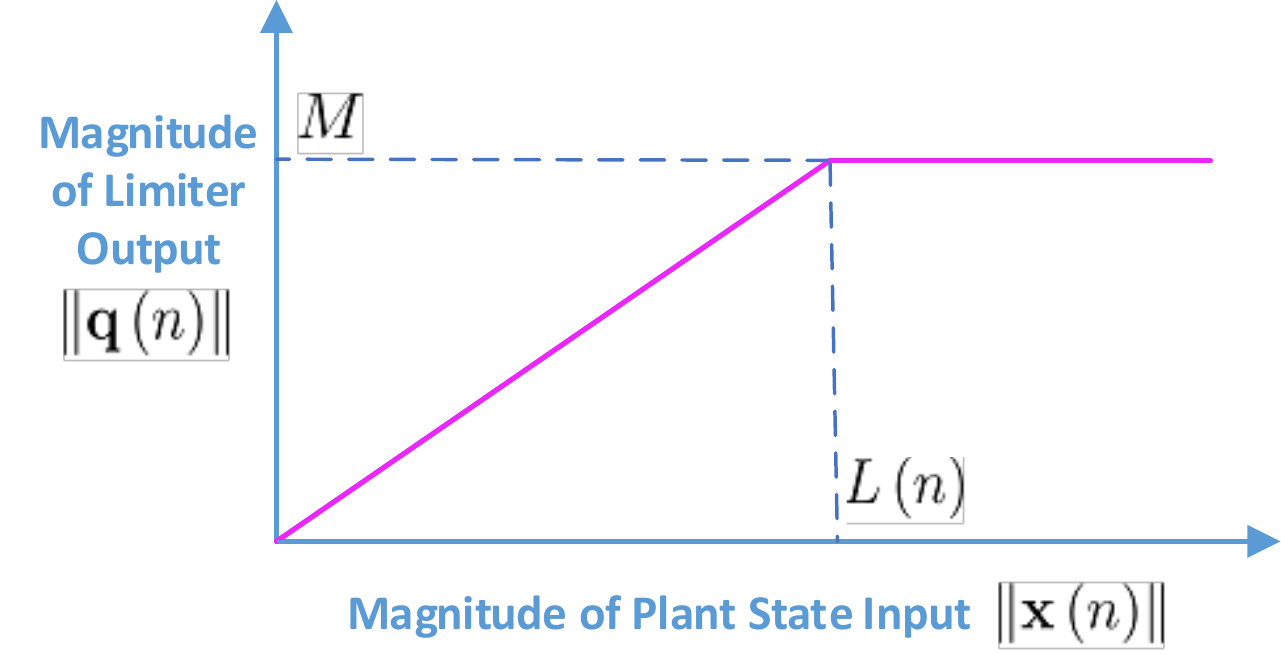}
\par\end{centering}

\caption{\label{fig: limiter-struc}Energy limiter at the sensor.}
\end{figure}

An energy limiter is a non-linear functional module, and the amplitude
relationship between its input and output is illustrated in Figure
\ref{fig: limiter-struc}. Specifically, an energy limiter is parameterized
by its dynamic range $L\left(n\right)$ and saturation value $M$.
If the limiter input signal's amplitude does not exceed the dynamic
range $L(n)$, then the input signals are allowed to pass with a scaling
of $\frac{M}{L(n)}$. On the hand, if the input signal's amplitude
is greater than the dynamic range $L(n)$, then its amplitude is attenuated
to the value $M$. The input of the energy limiter is the plant state
$\mathbf{x}(n)$ and the output is $\mathbf{q}(n)$, and the specific
structure of the energy limiter is given by: 
\begin{equation}
\mathbf{q}\left(n\right)=g\left(n\right)\mathbf{x}(n),\label{7equation-1}
\end{equation}
where $g\left(n\right)=\begin{cases}
\frac{M}{\left\Vert \mathbf{x}\left(n\right)\right\Vert }, & if\ \left\Vert \mathbf{x}\left(n\right)\right\Vert >L\left(n\right)\\
\frac{M}{L(n)}, & if\ \left\Vert \mathbf{x}\left(n\right)\right\Vert \leq L\left(n\right)
\end{cases}$ is a time varying constant.

The design of the limiter's dynamic range $L\left(n\right)$ is very
critical to the system performance. It's clear that, due to the nonlinearity
of the energy limiter, the magnitude information of $\mathbf{x}\left(n\right)$
will be partially lost when the limiter saturates, i.e., $\left\Vert \mathbf{x}\left(n\right)\right\Vert >L\left(n\right)$,
which will greatly deteriorate the state estimation quality at the
remote controller. We shall address how to design the limiter dynamic
range $L\left(n\right)$ in Section III.

The sensor then transmits the magnitude-limited plant state $\mathbf{q}\left(n\right)$
to the remote controller over a wireless MIMO fading channel, and
based on the limiter structure in (\ref{7equation-1}), we have $\mathbf{q}^{T}\left(n\right)\mathbf{q}\left(n\right)\leq M^{2}$.
Therefore, we restrict the energy availability constraint in (\ref{energyav})
as follows\footnote{It means that if the constraint in (\ref{asdsdsds-1}) is satisfied,
then the constraint in (\ref{energyav}) is satisfied. For analytical
tractability, we restrict the energy availability constraint (3) to
constraint (5) so that the dynamics of state estimation error covariance
is independent of the state $\mathbf{x}\left(n\right)$ and is therefore
analytically tractable. Since our proposed MIMO precoding $\mathbf{F}(n)$
has an event-driven structure (as shown in Theorem 1) and the sensor
can be in dormant mode most of the time to save energy, the difference
in terms of average energy consumption between constraint (5) and
constraint (3) is quite small.}: 
\begin{align}
M^{2}\mathrm{Tr}\left(\mathbf{F}^{H}(n)\mathbf{F}(n)\right)\tau\leq E(n).\label{asdsdsds-1}
\end{align}

\section{Dynamic MIMO Precoding via Lyapunov Optimization}

In this section, we shall address the design of the energy limiter,
and formulate the dynamic MIMO precoding problem to stabilize the
unstable MIMO dynamic plant using Lyapunov optimization. We first
have the following definition on the stability of the dynamic MIMO
plant \cite{neely2010stochasticlyaneely}:

\begin{definitn}
\textsl{(Stability of Dynamic MIMO Plant)}\label{definition-bounded-state}
The dynamic MIMO plant is stable if $\underset{N\rightarrow\infty}{\mathrm{limsup}}\frac{1}{N}\sum_{n=1}^{N}\mathbb{E}\left[\left\Vert \mathbf{x}\left(n\right)\right\Vert ^{2}\right]<\infty,$
where the expectation is taken with respect to the randomness of the
plant noise, the channel noise, the\textcolor{blue}{{} }MIMO channel
state, and the energy state.
\end{definitn}

\vspace{-0.3cm}

\subsection{Virtual State Estimation Covariance Process}

The nonlinear structure of the limiter introduces the nonlinear state
estimation, and thus the evolution of the plant state estimation mean
square error (MSE) is very difficult to be characterized explicitly.
To get around this, we first establish a tight analytical bound on
the state estimation MSE. Let $\hat{\mathbf{x}}\left(n\right)$ be
the minimum mean square error (MMSE) MIMO plant state estimate at
the controller. We have the following lemma, which gives an upper
bound of the MSE of the plant state estimation $\mathbb{E}\left[\left\Vert \mathbf{x}\left(n\right)-\hat{\mathbf{x}}\left(n\right)\right\Vert ^{2}\right]$

\begin{lemma}
\textsl{(State Estimation MSE Bound of the Stochastic Dynamic MIMO
Plant) }\label{lemma-mse-bound}For any control law $\mathbf{u}\left(n\right)$,
the plant state estimation MSE $\mathbb{E}\left[\left\Vert \mathbf{x}\left(n\right)-\hat{\mathbf{x}}\left(n\right)\right\Vert ^{2}\right]$
is bounded by a $\emph{virtual state estimation MSE process}$ according
to: 
\begin{align}
\mathbb{E}\left[\left\Vert \mathbf{x}\left(n\right)-\hat{\mathbf{x}}\left(n\right)\right\Vert ^{2}\right] & \leq\mathbb{E}\left[\mathrm{Tr}\left(\boldsymbol{\Sigma}\left(n\right)\right)\right],\label{eq:mse-bound}
\end{align}
where $\boldsymbol{\Sigma}\left(n\right)$ is a $\emph{virtual state estimation covariance process}$
with the following dynamics: 
\begin{align}
 & \boldsymbol{\Sigma}\left(n+1\right)=\mathbf{A}\big(\boldsymbol{\Sigma}\left(n\right)-\gamma\left(n\right)\boldsymbol{\Sigma}\left(n\right)\big(\mathbf{\mathbf{\mathbf{\tilde{F}}}}^{a}\left(n\right)\big)^{H}\Big(\mathbf{\mathbf{\mathbf{\tilde{F}}}}^{a}\left(n\right)\boldsymbol{\Sigma}\left(n\right)\nonumber \\
 & \hspace{1cm}\ \ \ \ \ \cdot\big(\mathbf{\mathbf{\mathbf{\tilde{F}}}}^{a}\left(n\right)\big)^{H}+\mathbf{I}\Big)^{-1}\mathbf{\mathbf{\mathbf{\tilde{F}}}}^{a}\left(n\right)\boldsymbol{\Sigma}\left(n\right)\big)\mathbf{A}^{T}+\mathbf{W},\label{dyna-sigma}
\end{align}
with initial value $\boldsymbol{\Sigma}\left(0\right)=\mathbf{0}$,
where $\gamma\left(n\right)=\mathbf{1}\left\{ \left\Vert \mathbf{x}\left(n\right)\right\Vert \leq L\left(n\right)\right\} $\textcolor{blue}{,
${\normalcolor {\normalcolor \mathbf{\tilde{F}}(n)=\mathbf{H}(n)\mathbf{F}(n)g(n})}$},
and $\mathbf{\mathbf{\mathbf{\tilde{F}}}}^{a}\left(n\right)=\begin{bmatrix}\mathbf{\mathbf{\tilde{F}}}\left(n\right)\\
\mathbf{\mathbf{\tilde{F}}}^{\dagger}\left(n\right)
\end{bmatrix}$ is an augmented $2N_{c}\times K$ matrix.\end{lemma}

\begin{IEEEproof}
Please see Appendix \ref{sub:Proof-of-mse-bound}.
\end{IEEEproof}

The upper bound of the MMSE in (6) can be achieved by using a low
complexity state estimation \footnote{Note that due to the non-linearity of the energy limiter, it is difficult
to compute the MMSE state estimator $\hat{\mathbf{x}}\left(n\right)$
.} $\widetilde{\hat{\mathbf{x}}}\left(n\right)=\mathbf{A}\widetilde{\hat{\mathbf{x}}}\left(n-1\right)+\gamma\left(n-1\right)\mathbf{A}\mathbf{K}\left(n-1\right)(\mathbf{y}^{a}\left(n-1\right)-\mathbf{\mathbf{\mathbf{\tilde{F}}}}^{a}\left(n-1\right)\widetilde{\hat{\mathbf{x}}}\left(n-1\right)),$
where $\mathbf{K}\left(n\right)=\boldsymbol{\Sigma}\left(n\right)(\mathbf{\mathbf{\mathbf{\tilde{F}}}}^{a}\left(n\right))^{H}(\mathbf{\mathbf{\mathbf{\tilde{F}}}}^{a}\left(n\right)\boldsymbol{\Sigma}\left(n\right)\big(\mathbf{\mathbf{\mathbf{\tilde{F}}}}^{a}\left(n\right)\big)^{H}+\mathbf{I})^{-1}$
and $\mathbf{y}^{a}\left(n\right)=\begin{bmatrix}\mathbf{\mathbf{y}}\left(n\right)\\
\mathbf{\mathbf{y}}^{\dagger}\left(n\right)
\end{bmatrix}$.

We consider a control law of the form $\mathbf{u}(n)=-\boldsymbol{\Psi}\mathbf{A}\widetilde{\hat{\mathbf{x}}}\left(n\right)$\footnote{Note that this form of control law is quite general and also covers
the certainty equivalent control which has been widely used in adaptive
control systems \cite{aastrom2013adaptiveadptivectrl}.} for the dynamic MIMO plant, where the feedback gain $\boldsymbol{\Psi}\mathbf{A}$
is chosen such that $\mathbf{A}-\mathbf{B}\boldsymbol{\Psi}\mathbf{A}$
is a Hurwitz matrix. Note that the certainty equivalent controller
is given by $\mathbf{u}(n)=-\boldsymbol{\Xi}\widetilde{\hat{\mathbf{x}}}\left(n\right)$,
where the feedback gain matrix is $\mathbf{\Xi}=\left(\mathbf{B}^{T}\mathbf{Z}\mathbf{B}+\mathbf{R}\right)^{-1}\mathbf{B}^{T}\mathbf{Z}$,
$\mathbf{Z}$ satisfies the following discrete time algebraic Ricatti
equation $\mathbf{Z}=\mathbf{A}^{T}\mathbf{Z}\mathbf{A}-\mathbf{A}^{T}\mathbf{Z}\mathbf{B}\left(\mathbf{B}^{T}\mathbf{Z}\mathbf{B}+\mathbf{R}\right)^{-1}\mathbf{B}^{T}\mathbf{Z}\mathbf{A}+\mathbf{P}$,
and $\mathbf{P}\in\mathbb{S}_{+}^{K}$ and $\mathbf{R}\in\mathbb{S}_{+}^{D}$
are the weighting matrices for the plant state deviation cost and
plant control cost of the LQG control associated with the certainty
equivalent controller \cite{CE1,CE2}. Hence, one possible way to
design such $\boldsymbol{\Psi}$ is let $\boldsymbol{\Psi}=-\left(\mathbf{B}^{T}\mathbf{Z}\mathbf{B}+\mathbf{R}\right)^{-1}\mathbf{B}^{T}\mathbf{Z}$.
From Lemma \ref{lemma-mse-bound}, if we can achieve stability in
the $\emph{virtual state estimation error covariance process}$, i.e.,
$\underset{N\rightarrow\infty}{\mathrm{limsup}}\sum_{n=1}^{N}{\color{blue}\frac{{\normalcolor 1}}{{\normalcolor N}}}\mathbb{E}\left[\mathrm{Tr}\left(\boldsymbol{\Sigma}\left(n\right)\right)\right]<\infty$,
then the actual plant state estimation MSE ($\mathbb{E}\left[\left\Vert \mathbf{x}\left(n\right)-\hat{\mathbf{x}}\left(n\right)\right\Vert ^{2}\right]$)
will also be bounded, which in turn leads to the bounded state process
as in Definition \ref{definition-bounded-state}. This is formally
stated in the following Lemma.

\begin{lemma}
\textsl{(Connection between Stability of }$\boldsymbol{\Sigma}\left(n\right)$
\textsl{and Stability of $\mathbf{x}\left(n\right)$)} \label{lemma-connnection}Under
the control law $\mathbf{u}(n)=-\boldsymbol{\Psi}\mathbf{A}\widetilde{\hat{\mathbf{x}}}\left(n\right)$,
if $\underset{N\rightarrow\infty}{\mathrm{limsup}}\frac{1}{N}\sum_{n=1}^{N}\mathbb{E}\left[\mathrm{Tr}\left(\boldsymbol{\Sigma}\left(n\right)\right)\right]<\infty$,
then $\underset{N\rightarrow\infty}{\mathrm{limsup}}\frac{1}{N}\sum_{n=1}^{N}\mathbb{E}\left[\left\Vert \mathbf{x}\left(n\right)\right\Vert ^{2}\right]<\infty.$ \end{lemma}

\begin{IEEEproof}
Please see Appendix \ref{sub:Proof-of-connection}.
\end{IEEEproof}

Note that the evolution of the virtual state estimation MSE process
$\boldsymbol{\Sigma}\left(n\right)$ depends on the dynamic MIMO precoder
$\mathbf{F}\left(n\right)$ according to (\ref{dyna-sigma}). As a
result, we will focus on the design of dynamic MIMO precoding $\mathbf{F}\left(n\right)$
in the MIMO NCS to achieve stability of $\boldsymbol{\Sigma}\left(n\right)$.
\vspace{-0.2cm}

\subsection{$\varepsilon$-Saturation Energy Limiter Design}

The magnitude information of the plant state is partially lost when
the limiter saturates. In conventional literature, bounded noise support
is assumed, and hence saturation can be avoided with proper dynamic
range design. In this work, we considered unbounded noise support,
and hence saturation cannot be completely eliminated. Instead, we
need to adapt the dynamic range $L\left(n\right)$ of the limiter
to maintain a small probability of saturation $\varepsilon$. We have
the following lemma on the design of the dynamic range of the limiter.

\begin{lemma}
\textsl{($\varepsilon$-Saturation Limiter Dynamic Range Adaptation)}
\label{lemma-limiter-range}Suppose the dynamic range $L\left(n\right)$
of the limiter evolves according to the following: 
\begin{align}
L\left(n\right)= & \frac{1}{\sqrt{\varepsilon}}\left(1+\left\Vert \mathbf{A}-\mathbf{B}\boldsymbol{\Psi}\mathbf{A}\right\Vert \Theta\right)\nonumber \\
 & \cdot\left(\left\Vert \mathbf{B}\boldsymbol{\Psi}\mathbf{A}\right\Vert \sqrt{\mathrm{Tr}\left(\boldsymbol{\Sigma}\left(n\right)\right)-\mathrm{Tr}\left(\mathbf{W}\right)}+\sqrt{\mathrm{Tr}\left(\mathbf{W}\right)}\right),\label{eq:L(n)}
\end{align}
where
\begin{align}
\Theta & =\frac{1}{\mu_{min}\left(\mathbf{T}\right)}\bigg(\left\Vert \left(\mathbf{A}-\mathbf{B}\boldsymbol{\Psi}\mathbf{A}\right)^{T}\mathbf{Q}\right\Vert \nonumber \\
 & +\left(\left\Vert \left(\mathbf{A}-\mathbf{B}\boldsymbol{\Psi}\mathbf{A}\right)^{T}\mathbf{Q}\right\Vert ^{2}+\mu_{min}\left(\mathbf{T}\right)\left\Vert \mathbf{Q}\right\Vert \right)^{\nicefrac{1}{2}}\bigg)
\end{align}
is a constant ($\mathbf{Q}$ and $\mathbf{T}$ are any positive definite
symmetric matrices such that $\left(\mathbf{A}-\mathbf{B}\boldsymbol{\Psi}\mathbf{A}\right)^{T}\mathbf{Q}\left(\mathbf{A}-\mathbf{B}\boldsymbol{\Psi}\mathbf{A}\right)-\mathbf{Q}=-\mathbf{T}$).
Then 
\begin{align}
\mathrm{Pr}\left(\left\Vert \mathbf{x}\left(n\right)\right\Vert >L\left(n\right)\right) & \leq\varepsilon
\end{align}
holds for any time slot $n$. \end{lemma}

\begin{IEEEproof}
Please see Appendix \ref{sub:Proof-of-limiter-range}.
\end{IEEEproof}

\begin{remrk}
\textsl{(Computation of $L\left(n\right)$ at the Sensor) }Since the
input state process $\mathbf{x}\left(n\right)$ is a non-stationary
random process, in order to maintain a small saturation probability
$\varepsilon$, the dynamic range $L\left(n\right)$ of the limiter
needs to be adaptive as in (\ref{eq:L(n)}) to keep track of the instantaneous
covariance of the input state process $\mathbf{x}\left(n\right)$.
Note that the limiter's dynamic range $L\left(n\right)$ is a function
of the virtual state estimation covariance $\boldsymbol{\Sigma}\left(n\right)$.
The sensor can obtain $\boldsymbol{\Sigma}\left(n\right)$ via the
dynamics (\ref{dyna-sigma}) based on the local information only without
explicit signaling feedback, and hence it is implementation friendly.
Due to the local availability of $\boldsymbol{\Sigma}\left(n\right)$
and $L\left(n\right)$ at the sensor, the MIMO precoding solution
in Theorem 1 can be implemented\textcolor{blue}{{} }at the sensor in
a decentralized way.\vspace{-0.2cm}

\end{remrk}

\subsection{Dynamic MIMO Linear Precoding via Lyapunov Optimization}

\vspace{-0.2cm}

We focus on deriving a dynamic MIMO precoding policy to achieve stability
of the virtual covariance process $\boldsymbol{\Sigma}\left(n\right)$
using Lyapunov techniques \cite{neely2005dynamiclyaopt,neely2006energy}.
Intuitively, a negative term in the Lyapunov drift is a stabilizing
force that pulls the system state back to the equilibrium point. As
a result, we shall design the dynamic MIMO precoder to maximize the
negative Lyapunov drift to achieve stability under the renewable energy
resource constraints. 

We define a Lyapunov function as follows \cite{neely2006energy}:
\begin{equation}
V\left(n\right)=\frac{1}{2}\mathrm{Tr}\left(\boldsymbol{\Sigma}\left(n\right)\right)+\frac{1}{2}\left(E\left(n\right)-\theta\right)^{2}.\label{eq: drift}
\end{equation}
The associated Lyapunov drift is given by 
\begin{align}
 & \Lambda\left(\boldsymbol{\Sigma}\left(n\right)\right)\nonumber \\
 & =\mathbb{E}\left\{ \left.V\left(n+1\right)-V\left(n\right)\right|\boldsymbol{\Sigma}\left(n\right)\right\} ,\label{eq:one step conditional drift}
\end{align}
where the expectation $\mathbb{E}$ is w.r.t. the randomness of the
channel state and the energy state under a given MIMO precoding rule
$\mathbf{F}(n)$. We have the following lemma on the Lyapunov drift.

\begin{lemma}
\textsl{(Lyapunov Drift)} \label{lemma- conditional drift}Given a
MIMO precoding rule $\mathbf{F}(n)$, the Lyapunov drift can be upper
bounded as follows:
\begin{align}
 & \Lambda\left(\boldsymbol{\Sigma}\left(n\right)\right)\leq\frac{1}{2}\mathrm{Tr}(\mathbf{W})+\mathbb{E}\left[\left(\alpha\left(n\right)+\theta\right)^{2}\right]\nonumber \\
 & +\mathbb{E}\bigg\{\underbrace{\frac{1}{2}\left\Vert \mathbf{A}\mathbf{A}^{T}\right\Vert \bigg[\varepsilon\mathrm{Tr}\left(\boldsymbol{\Sigma}\left(n\right)\right)}_{"bad"\ term\ causing\ instability}+\nonumber \\
 & \underbrace{\mathrm{Tr}\left(2\frac{M^{2}}{L^{2}(n)}\mathrm{Re}\left\{ \mathbf{F}^{H}\left(n\right)\mathbf{H}^{H}(n)\mathbf{H}(n)\mathbf{F}\left(n\right)\right\} +\boldsymbol{\Sigma}\left(n\right)^{-1}\right)^{-1}}_{"bad"\ term\ causing\ instability}\bigg]\nonumber \\
 & +\underbrace{M^{2}\mathrm{Tr}\left(\mathbf{F}^{H}(n)\mathbf{F}(n)\right)\tau\left(\theta-E\left(n\right)\right)}_{"bad"\ term\ causing\ instability}\nonumber \\
 & -\underbrace{\frac{1}{2}\mathrm{Tr}\left(\boldsymbol{\Sigma}\left(n\right)\right)}_{\underset{(stabilizing\ force)}{"good"\ term}}\bigg\}\bigg|\boldsymbol{\Sigma}\left(n\right)\bigg\}.\label{eq:lya-drift}
\end{align}
\end{lemma}

\begin{IEEEproof}
Please see Appendix \ref{sub:Proof-of-pdf-property-2-1-1}.
\end{IEEEproof}

The first expectation on the R.H.S of (\ref{eq:lya-drift}) is taken
w.r.t. the randomness of energy arrival, the second expectation on
the R.H.S of (\ref{eq:lya-drift}) is taken w.r.t. the randomness
of both the channel state and the energy state under given $\boldsymbol{\Sigma}\left(n\right)$
and MIMO precoding rule $\mathbf{F}(n)$. Note that the negative Lyapunov
drift plays a central role when applying the Lyapunov drift theory
to analyze the stability of dynamic systems. Intuitively, the negative
Lyapunov drift is a stabilizing force that pulls the system state
back to the equilibrium point. The drift term in equation (\ref{eq:lya-drift})
that contributes to positive Lyapunov drift is bracketed as a ``bad''
term for stabilization of $\boldsymbol{\Sigma}\left(n\right)$, because
positive Lyapunov drift leads to instability of $\boldsymbol{\Sigma}\left(n\right)$.
On the contrary, the drift term that contributes to negative Lyapunov
drift is bracketed as a ``good'' term for stabilization of $\boldsymbol{\Sigma}\left(n\right)$,
because negative Lyapunov drift is the stabilizing force for the stability
of $\boldsymbol{\Sigma}\left(n\right)$. Hence, we focus on deriving
a dynamic MIMO precoder to minimize the drift (\ref{eq:lya-drift}).
This is equivalent to considering the following optimization problem:

\begin{problem}
\textit{(MIMO Precoding via Lyapunov Optimization)} \label{problem-lya-opt}For
given realizations $E(n)=E$, $\boldsymbol{\Sigma}\left(n\right)=\boldsymbol{\Sigma}$,
and $\mathbf{H}(n)=\mathbf{H}$, the MIMO precoding $\mathbf{F}\left(n\right)$
is given by the solution of the following problem: 
\begin{align*}
 & \underset{\mathrm{\mathbf{F}}\left(n\right)}{\mathrm{min}}\ \ M^{2}\mathrm{Tr}\left(\mathbf{F}^{H}(n)\mathbf{F}(n)\right)\tau\left(\theta-E\right)\\
 & +\frac{\left\Vert \mathbf{A}\mathbf{A}^{T}\right\Vert }{2}\mathrm{Tr}\left(2\frac{M^{2}}{L^{2}(\boldsymbol{\Sigma})}\mathrm{Re}\left\{ \mathbf{F}^{H}(n)\mathbf{H}^{H}\mathbf{H}\mathbf{F}(n)\right\} +\boldsymbol{\Sigma}^{-1}\right)^{-1}\\
 & s.t.\ \ M^{2}\mathrm{Tr}\left(\mathbf{F}^{H}(n)\mathbf{F}(n)\right)\tau\leq E,
\end{align*}
where $L(\boldsymbol{\Sigma})$ is given in (\ref{eq:L(n)}).
\end{problem}

\vspace{-0.2cm}

\begin{center}
\fbox{\begin{minipage}[t]{1\columnwidth}%
Challenge 2: Closed-form Dynamic MIMO Precoder Design.%
\end{minipage}}
\par\end{center}

While Problem 1 is equivalent to a convex problem, it is still very
challenging to obtain closed-form solution. This is due to the tight
coupling in the objective function (involving a mixture of real, trace
and inverse).

\section{Event-Driven Energy Harvesting and MIMO Precoding Solution}

In this section, we shall give the MIMO precoding solution to Problem
1 and discuss its structural properties.

\subsection{Drift Minimizing MIMO Precoder}

The underlying structure of the objective function of Problem \ref{problem-lya-opt}
provides some key insights into the MIMO precoding solution structure.
Specifically, if $\mathrm{Re}\left\{ \mathbf{F}^{H}(n)\mathbf{H}^{H}\mathbf{H}\mathbf{F}(n)\right\} $,
$\boldsymbol{\Sigma}^{-1}$ and $\mathbf{F}^{H}(n)\mathbf{F}(n)$
can be simultaneously diagnalized, then the objective function is
substantially simplified because the matrix operations will only involve
diagonal matrices. This may be feasible if $\mathbf{F}(n)$ can be
expressed as a product of three matrices, where the leftmost matrix
is the Hermitian of the right singular matrix of $\mathbf{H}$ and
the rightmost matrix is the Hermitian of the left right singular matrix
of $\boldsymbol{\Sigma}$. Inspired by this observation, let the singular
value decomposition of the matrix of $\mathbf{H}$ be $\mathbf{H}=\mathbf{V}\boldsymbol{\Pi}\mathbf{U}^{H}$,
where\textcolor{blue}{{} ${\normalcolor \mathbf{V}\in\mathbb{C}^{N_{c}\times N_{c}}}$
}and $\mathbf{U}\in\mathbb{C}^{N_{s}\times N_{s}}$ are unitary matrices,
and $\boldsymbol{\Pi}$ is a rectangular diagonal matrix with diagonal
elements in a descending order. Denote $\boldsymbol{\Pi}_{K}$ as
the leading principal minor of order $K$ of $\boldsymbol{\Pi}$.
Let the eigenvalue decomposition of $\boldsymbol{\Sigma}^{-1}$ be
$\boldsymbol{\Sigma}=\mathbf{S}\boldsymbol{\Lambda}\mathbf{S}^{T}$,
where $\mathbf{S}\in\mathbb{C}^{K\times K}$ is an orthogonal matrix,
and $\boldsymbol{\Lambda}\in\mathbb{\mathbb{R}}^{K\times K}$ is diagonal
with diagonal elements in a descending order. Then the $\emph{drift-minimizing}$
solution of Problem 1 is given as follows.

\begin{thm}
\textsl{(Drift-Minimizing Solution)}\label{solution} The $\emph{drift-minimizing}$
MIMO precoding is as follows:
\begin{itemize}
\item \textbf{Dormant Mode:} If $\theta\mathbf{I}-\big(\frac{\left\Vert \mathbf{A}\mathbf{A}^{T}\right\Vert \boldsymbol{\Lambda}^{2}\boldsymbol{\Pi}_{K}^{2}}{\tau L^{2}(\boldsymbol{\Sigma})}+E\mathbf{I}\big)$
is positive definite, then $\mathbf{F}^{\ast}\left(n\right)=\mathbf{0}.$ 
\item \textbf{Active Mode: }If $\theta\mathbf{I}-\big(\frac{\left\Vert \mathbf{A}\mathbf{A}^{T}\right\Vert \boldsymbol{\Lambda}^{2}\boldsymbol{\Pi}_{K}^{2}}{\tau L^{2}(\boldsymbol{\Sigma})}+E\mathbf{I}\big)$
is not positive definite, then 
\begin{align*}
 & \mathbf{F}^{\ast}\left(n\right)=\frac{L(\boldsymbol{\Sigma})}{M}\mathbf{U}\\
 & \cdot\bigg[\begin{array}{c}
\boldsymbol{\Pi}_{K}^{-1}\Big(\frac{1}{2}\Big[\frac{\boldsymbol{\Pi}_{K}}{L\left(\boldsymbol{\Sigma}\right)}\sqrt{\frac{\left\Vert \mathbf{A}\mathbf{A}^{T}\right\Vert }{\left(\left[\theta-E\right]^{+}+\beta\right)\tau}}-\boldsymbol{\Lambda}^{-1}\Big]^{+}\Big)^{\nicefrac{1}{2}}\mathbf{S}^{T}\\
\mathbf{0}
\end{array}\bigg],
\end{align*}
 where $\beta$ is given by (\ref{eq:solution-struc}). 
\begin{figure*}[tbh]
\begin{equation}
\beta=\begin{cases}
chosen\ such\ that\  & if\ \mathrm{Tr}\Big(\frac{1}{2}\boldsymbol{\Pi}_{K}^{-2}\Big[\frac{\boldsymbol{\Pi}_{K}}{L\left(\boldsymbol{\Sigma}\right)}\sqrt{\frac{\left\Vert \mathbf{A}\mathbf{A}^{T}\right\Vert }{\left(\theta-E\right)\tau}}-\boldsymbol{\Lambda}^{-1}\Big]^{+}\Big)\geq\frac{E}{\tau L^{2}(\boldsymbol{\Sigma})};\\
M^{2}\mathrm{Tr}\left({\color{blue}{\normalcolor \left(\mathbf{F}^{\ast}(n)\right)^{H}\mathbf{F}^{\ast}(n)}}\right)\tau=E,\\
\\
0, & otherwise.
\end{cases}\label{eq:solution-struc}
\end{equation}

\hrulefill

\end{figure*}
 
\end{itemize}
\end{thm}

\begin{IEEEproof}
Please see Appendix D.
\end{IEEEproof}

\begin{remrk}
\textsl{(Interpretation on the Structure of the Drift-minimizing MIMO
Precoding Solution)  }
\end{remrk}

\textit{Event-driven Structure:} The MIMO precoding solution in Theorem
\ref{solution} also has an \emph{event-driven structure}, in the
sense that the sensor either transmits or shuts down $\emph{aperiodically}$
depending only on whether the dynamic threshold $\theta\mathbf{I}-\big(\frac{\left\Vert \mathbf{A}\mathbf{A}^{T}\right\Vert \boldsymbol{\Lambda}^{2}\boldsymbol{\Pi}_{K}^{2}}{\tau L^{2}(\boldsymbol{\Sigma})}+E\mathbf{I}\big)$
is positive definite or not.

\begin{itemize}
\item Large $\|\boldsymbol{\Pi}_{K}\|$ (good channel condition) or large
$E$ (sufficient energy in the storage) leads to active mode of the
sensor, which means it is better that the sensor be active when there
are good transmission opportunities. 
\item Large MIMO plant state estimation error implies large $\|\boldsymbol{\Sigma}\|$,
as shown in Lemma \ref{lemma-mse-bound}, which also leads to the
active mode of the sensor. This is because a large MIMO plant state
estimation error means high transmission urgency and hence it is better
that the sensor be active.
\end{itemize}

\begin{figure}[tbh]
\begin{centering}
\includegraphics[clip,scale=0.6]{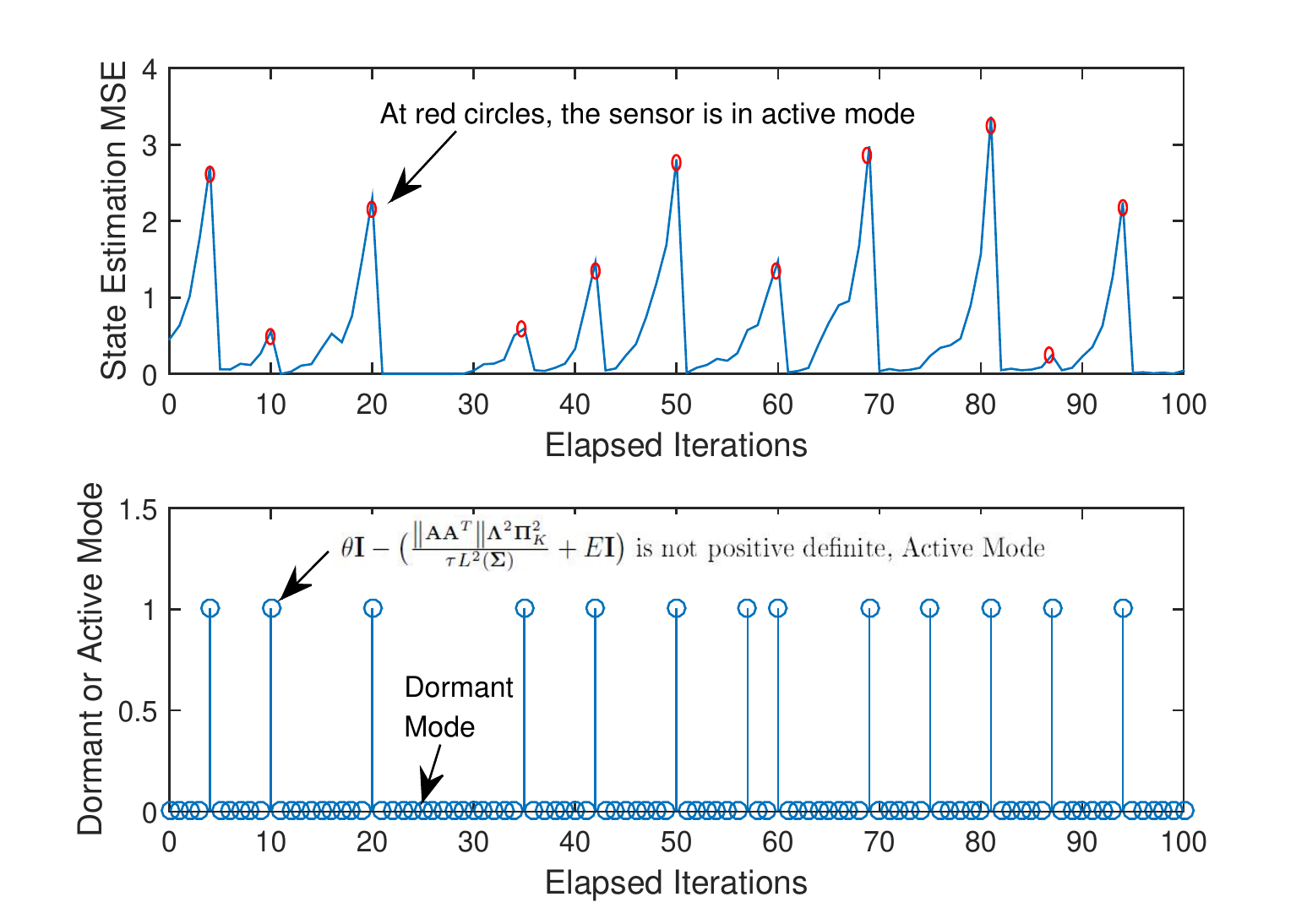}
\par\end{centering}

\caption{\label{fig: threshold-evolution}Sample path of state estimation error
and transitions between active and dormant modes. The system parameters
are configured as follows:$\mathbf{A}=\left(\protect\begin{array}{cc}
1.3 & 0.1\protect\\
-0.2 & 1.2
\protect\end{array}\right)$, $\mathbf{\mathbf{B}}=\mathrm{diag(1,1)}$, $\mathbf{\mathbf{W}}=\mathrm{diag(1,2)}$,
$\mathbf{\mathbf{Q}}=\mathrm{diag(1,1)}$, $\mathbf{\mathbf{\Psi}}=\mathrm{diag(0.25,0.25)}$,
$\varepsilon=0.01$, $M=1$, $N_{s}=3$, $N_{c}=2$, $\tau=0.01$,
$\mathbb{E}\left[\alpha\right]=5$ and $\theta=30.$}
\end{figure}

Figure \ref{fig: threshold-evolution} illustrates a sample path of
the state estimation error and transition between the active and dormant
modes. It can be observed that the state estimation error \textcolor{blue}{${\normalcolor \left\Vert \mathbf{x}\left(n\right)-\hat{\mathbf{x}}\left(n\right)\right\Vert ^{2}}$}
increases during the dormant modes and is reset to a low value during
the active modes. As such, the MIMO precoding solution has an \textit{event-driven
}structure with $\emph{aperiodic\ reset}$ of state estimation error.

Note that the virtual state estimation covariance $\boldsymbol{\Sigma}$
can be decomposed as $\boldsymbol{\Sigma}=\mathbf{S}\boldsymbol{\Lambda}\mathbf{S}^{H}$
with $\mathbf{S}$ and $\boldsymbol{\Lambda}$ being unitary and diagonal,
respectively. Hence, each diagonal element of $\boldsymbol{\Lambda}$
corresponds to a subsystem with $\boldsymbol{\Lambda}_{ii}$ as the
state estimation error of the $i$-th subsystem.

\textit{Dynamic Spatial Channel Activation Structure: }The MIMO precoding
solution $\mathbf{F}^{\ast}\left(n\right)$ dynamically activates
the spatial channels. Specifically, whether the $i$-th spatial channel
(the spatial channel corresponds to the $i$-th subsystem) is activated
or not depends only on whether the dynamic threshold $\frac{\left\Vert \mathbf{A}\mathbf{A}^{T}\right\Vert \left(\boldsymbol{\Lambda}\boldsymbol{\Pi}_{K}\right)_{ii}^{2}}{\tau L^{2}(\boldsymbol{\Sigma})}-\left[\theta-E\right]^{+}-\beta$
is positive or not. 

\textit{Eigenvalue Water-filling Structure:} The MIMO precoding solution
$\mathbf{F}^{\ast}\left(n\right)$ also has an $\emph{eigenvalue water-filling structure}$.
Specifically, for the $i$-th subsystem, $\frac{\left(\boldsymbol{\Pi}_{K}\right)_{ii}}{L\left(\boldsymbol{\Sigma}\right)}\sqrt{\frac{\left\Vert \mathbf{A}\mathbf{A}^{T}\right\Vert }{\left(\left[\theta-E\right]^{+}+\beta\right)\tau}}$
is the dynamic water level which adapts to the virtual state estimation
covariance $\boldsymbol{\Sigma}$, energy storage $E$, and the $i$-th
spatial channel $\left(\boldsymbol{\Pi}_{K}\right)_{ii}$, and $\left(\boldsymbol{\Lambda}\right)_{ii}^{-1}$
is the dynamic seabed level. Large energy storage (i.e., large $E$)
or better channel condition of the $i$-th spatial channel (i.e.,
large $\left(\boldsymbol{\Pi}_{K}\right)_{ii}$) will lead to a high
water level which will allow the sensor to allocate more transmission
energy for the $i$-th subsystem. High transmission urgency (i.e.,
large $\left(\boldsymbol{\Lambda}\right)_{ii}$) will lead to a low
seabed level, which induces more energy to be allocated to the subsystem.

\begin{figure}[tbh]
\begin{centering}
\subfloat[System model of the noisy decoupled plant system.]{\includegraphics[clip,width=0.9\columnwidth]{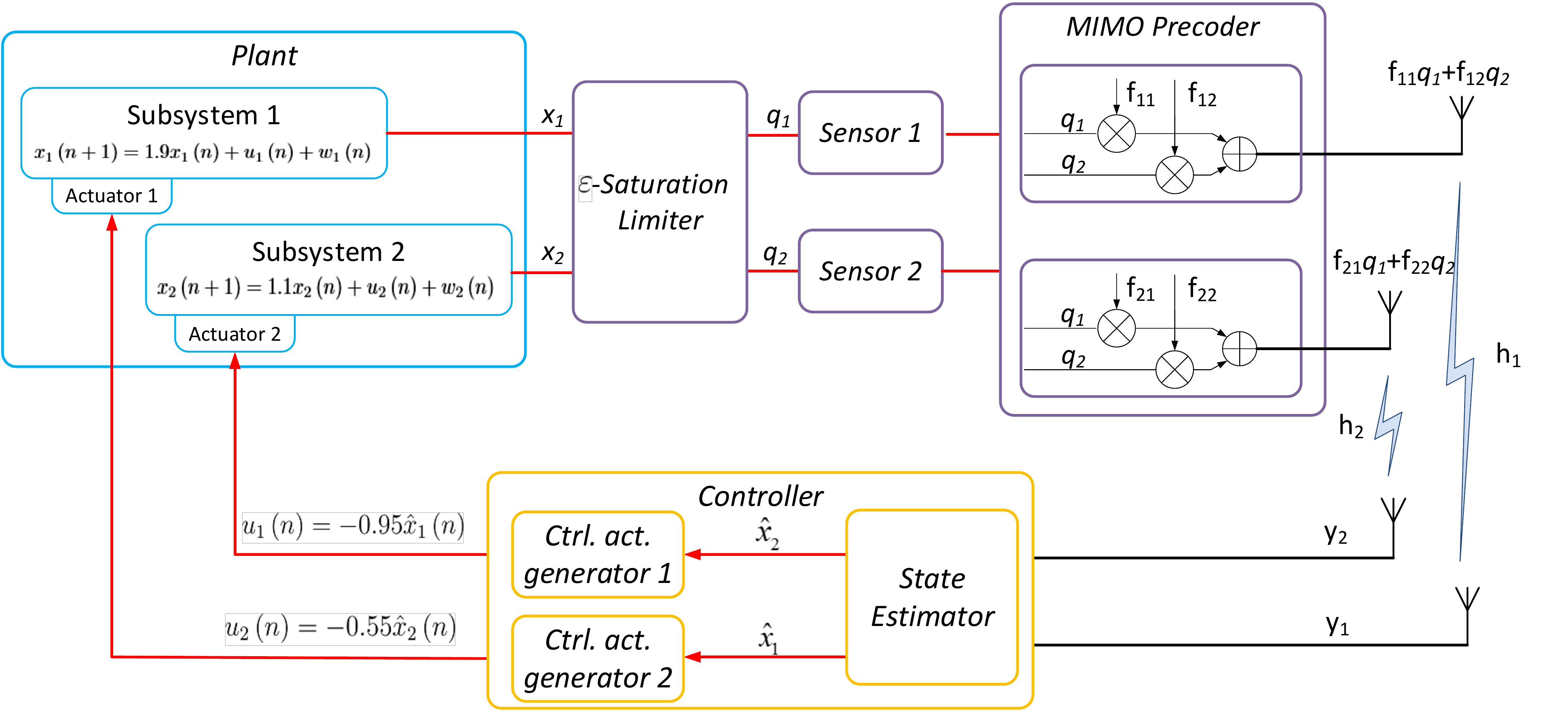}}
\par\end{centering}

\begin{centering}
\subfloat[Eigenvalue water-filling structure.]{\includegraphics[clip,width=0.9\columnwidth]{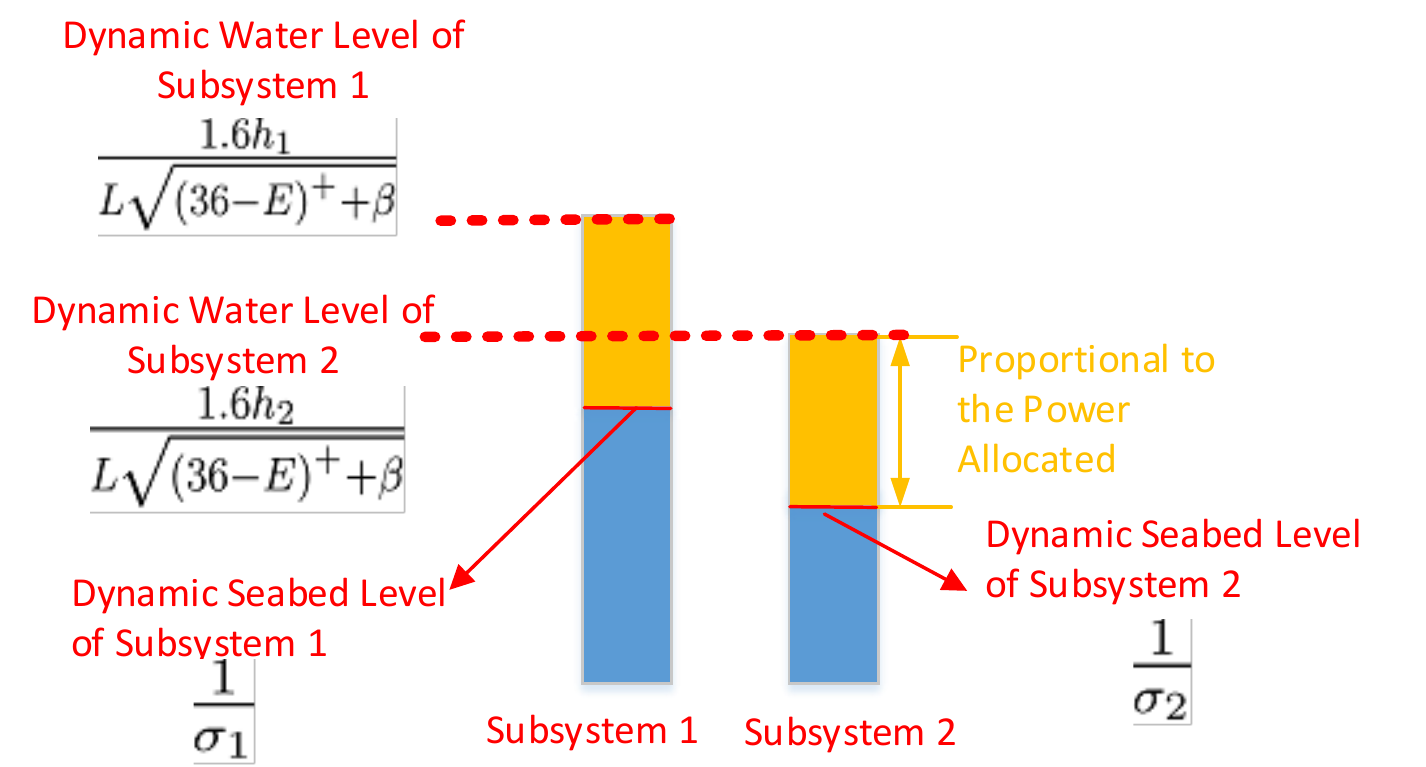}

}
\par\end{centering}

\caption{\label{fig: example-figure}Illustrations of the system model and
the eigenvalue water-filling structure of the\textcolor{blue}{{} }drift-minimizing
MIMO precoding solution in Example 1\textcolor{blue}{.}}
\end{figure}

\begin{remrk}
\textsl{(Comparison with Traditional MIMO Precoding Solutions)  }
\end{remrk}

The traditional waterfilling MIMO precoding solution \cite{epi-lemma1}
that maximizing the physical layer capacity has the form $\mathbf{F}\left(n\right)=\mathbf{U}\mathbf{\Upsilon}$,
where $\left(\mathbf{\Upsilon}\right)_{ii}=\left[\gamma-\frac{1}{\left(\boldsymbol{\Pi}_{K}\right)_{ii}}\right]^{+}$
for all $i=1,\ldots,K$, the water level $\gamma$ is chosen such
that $\sum_{i=i}^{K}\left(\mathbf{\Upsilon}\right)_{ii}=\frac{E\left(n\right)}{M^{2}\tau}$,
and the other elements of $\mathbf{\Upsilon}$ are zero. The traditional
waterfilling solution is only adapted to the channel state. The waterfilling
solution under energy harvesting constraints in \cite{EH-precoder}
maximizes the physical layer capacity subject to the energy causality
constraints, and the solution has the form $\mathbf{F}\left(n\right)=\mathbf{U}\mathbf{\Upsilon}$,
where $\left(\mathbf{\Upsilon}\right)_{ii}=\left[\frac{1}{\sum_{j=i}^{K+1}\lambda_{j}-\sum_{j=i}^{K}\mu_{j}}-\frac{1}{\left(\boldsymbol{\Pi}_{K}\right)_{ii}}\right]^{+}$
for all $i=1,\ldots,K$, $\lambda_{j}$ are the Lagrange multiplier
that enforce energy causality and $\mu_{j}$ are the Lagrange multipliers
that enforce no-energy overflow conditions. Similarly, the waterfilling
solution under energy harvesting constraints also only adapts to the
energy state and channel state. In contrast, the proposed MIMO precoding
solution is truly dynamic and adaptive to the MIMO channel state,
sensor energy state and plant state transmission urgency. It is noticed
that the traditional waterfilling solution and the waterfilling solution
with energy harvesting constraints both fail to exploit plant state
transmission urgency and therefore, they have inferior performance
as shown in Section VI.

\subsection{Application Example}

We consider the following simple toy example to illustrate the structural
properties of the MIMO precoding solution.

\begin{example}
\textsl{(Noisy Decoupled Plant System)} We consider an aggregation
of two decoupled SISO plant subsystems as illustrated in Figure \ref{fig: example-figure}.
The dynamics of the two subsystems are given by:
\begin{align*}
 & x_{1}\left(n+1\right)=1.6x_{1}\left(n\right)+u_{1}\left(n\right)+w_{1}\left(n\right)\ \ \ (Subsystem\ 1)\\
 & x_{2}\left(n+1\right)=1.1x_{2}\left(n\right)+u_{2}\left(n\right)+w_{2}\left(n\right)\ \ \ (Subsystem\ 2),
\end{align*}
where $w_{1}\left(n\right)$ and $w_{2}\left(n\right)$ are independent
Gaussian random noise with zero mean and covariance 1. Let the control
rule be $u_{1}\left(n\right)=-0.8\hat{x}_{1}\left(n\right)$ and $u_{2}\left(n\right)=-0.55\hat{x}_{2}\left(n\right)$.
Let $\theta=36$, $\tau=1$ and the parameters of the $\varepsilon$-saturation
limiter be $\varepsilon=0.1$, $M=1$ and $\mathbf{\mathbf{Q}}=\mathbf{I}$.
The energy-harvesting sensor observes the limiter output $\mathbf{q}\left(n\right)=\left(q_{1}\left(n\right),q_{2}\left(n\right)\right)$,
applies a precoder $\mathbf{F}=\left[\begin{array}{cc}
f_{11} & f_{12}\\
f_{21} & f_{22}
\end{array}\right]$ and sends the magnitude-limited plant state to the controllers via
a decoupled parallel channel $\mathbf{H}=\left(\begin{array}{cc}
h_{1} & 0\\
0 & h_{2}
\end{array}\right)$. This corresponds to a special case of the NCS system with $K=2$,
$N_{s}=2$, $N_{c}=2$, $\mathbf{A}=\left(\begin{array}{cc}
1.6 & 0\\
0 & 1.1
\end{array}\right)$, $\mathbf{B}=\mathbf{\mathbf{W}}=\mathbf{I}$, and $\boldsymbol{\Psi}=\frac{1}{2}\mathbf{I}$. 
\end{example}

According to Theorem \ref{solution}, the MIMO precoding solution
for Example 1 is given by:
\begin{align*}
\mathbf{F}^{\ast}\left(n\right)= & \frac{L}{\sqrt{2}}\mathrm{diag}\bigg\{\frac{1}{\left|h_{1}\right|}\sqrt{\left[\frac{1.6\left|h_{1}\right|}{L\sqrt{\left(36-E\right)^{+}+\beta}}-\frac{1}{\sigma_{1}}\right]^{+}}\\
 & ,\frac{1}{\left|h_{2}\right|}\sqrt{\left[\frac{1.6\left|h_{2}\right|}{L\sqrt{\left(36-E\right)^{+}+\beta}}-\frac{1}{\sigma_{2}}\right]^{+}}\bigg\},
\end{align*}
 where $\boldsymbol{\Sigma}_{n}=\left[\begin{array}{cc}
\sigma_{1} & 0\\
0 & \sigma_{2}
\end{array}\right]$ is the virtual state estimation covariance matrix, $L=22.36+7.91\sqrt{\sigma_{1}+\sigma_{2}-2}$,
and $\beta=0$ if $\left\Vert \mathbf{F}^{\ast}\left(n\right)\right\Vert _{F}^{2}<E$,
else $\beta$ is chosen such that $\left\Vert \mathbf{F}^{\ast}\left(n\right)\right\Vert _{F}^{2}=E$.

\begin{itemize}
\item \textbf{When the Sensor has to be Active?} Based on the \textit{event-driven
structure} in Remark 3, the sensor will be activated if either $36-E-\frac{2.56h_{1}^{2}\sigma_{1}^{2}}{\left(22.36+7.91\sqrt{\sigma_{1}+\sigma_{2}-2}\right)^{2}}$
or $36-E-\frac{2.56h_{2}^{2}\sigma_{2}^{2}}{\left(22.36+7.91\sqrt{\sigma_{1}+\sigma_{2}-2}\right)^{2}}$
is negative. Therefore, better channel condition (large $h_{1}$ or
$h_{2}$), large energy storage $E$ or increased transmission urgency
(large $\sigma_{1}$ or $\sigma_{2}$) will lead to the active mode
of the sensor\textcolor{blue}{. }
\end{itemize}

\begin{figure}[tbh]
\begin{centering}
\subfloat[Decision Region of Spatial Channel Activation w.r.t. $h_{2}$ and
$\sigma_{2}$ with $h_{1}=4$, $\sigma_{1}=70$ and $E=12$.]{\includegraphics[clip,width=0.8\columnwidth]{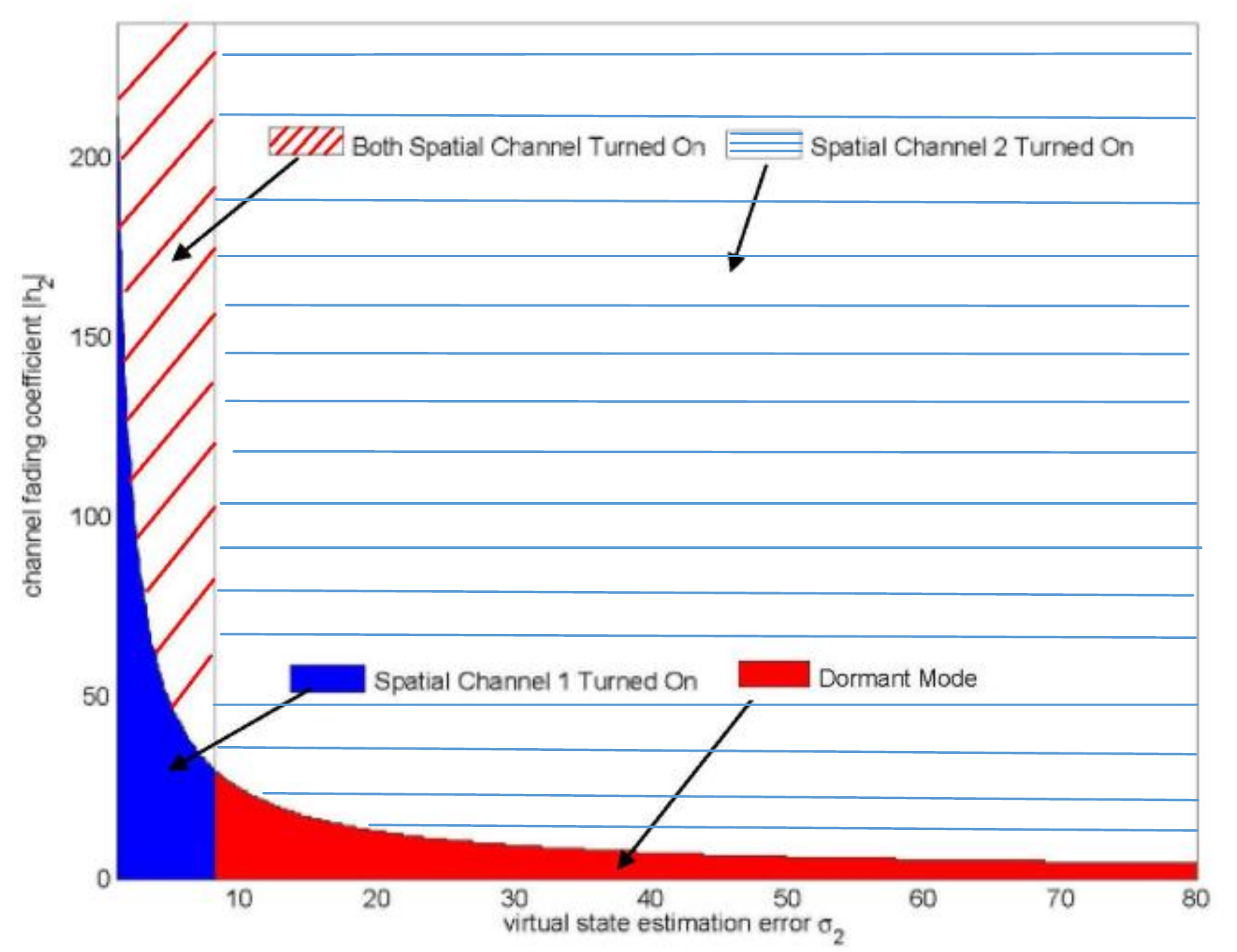}}
\par\end{centering}

\begin{centering}
\subfloat[Decision Region of Spatial Channel Activation w.r.t. $h_{2}$ and
$\sigma_{2}$ with $h_{1}=4$, $\sigma_{1}=70$ and $E=20$.]{\includegraphics[clip,width=0.8\columnwidth]{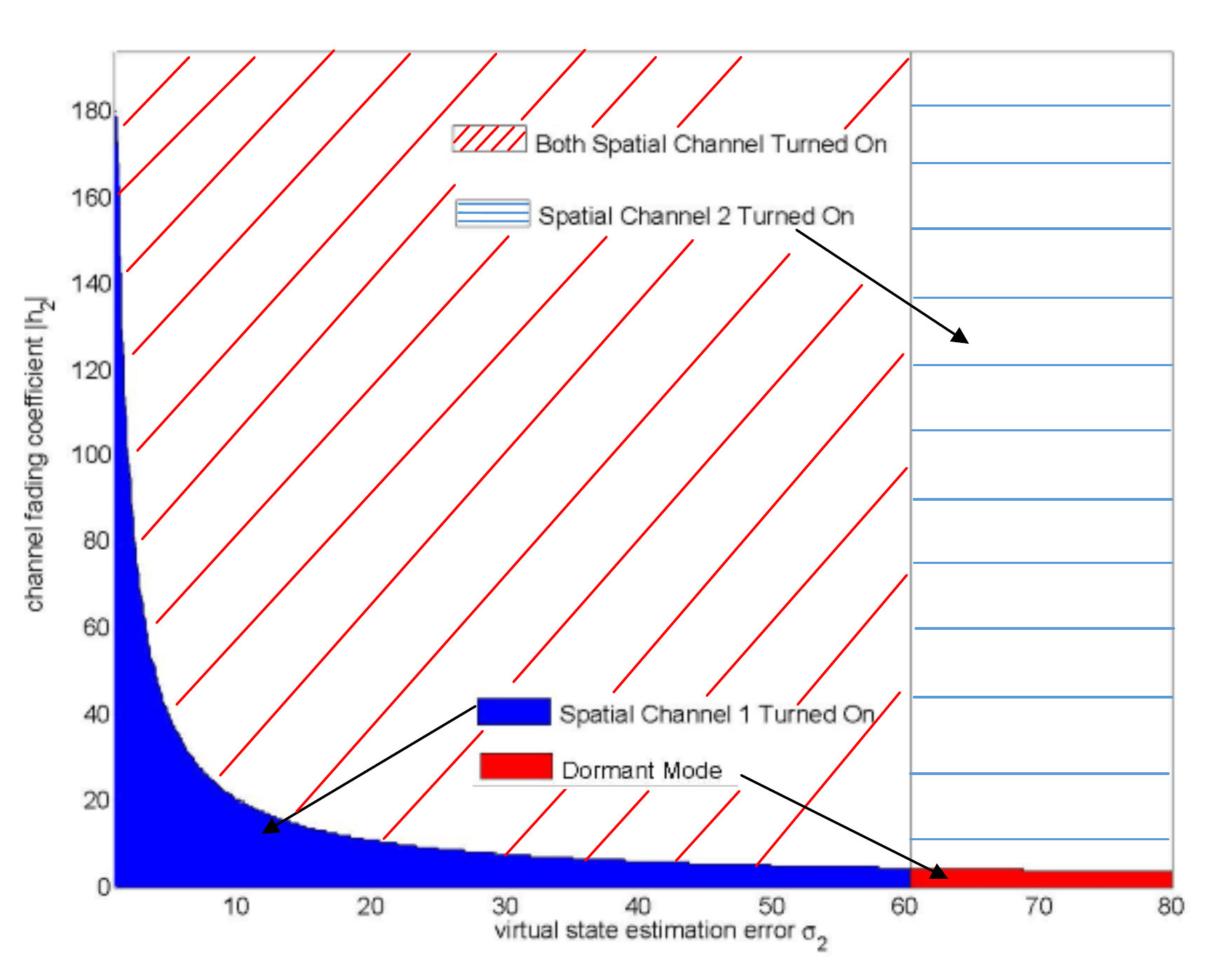}}
\par\end{centering}

\caption{\label{fig: example-figure-1}Decision Region of Spatial Channel Activation
w.r.t. $h_{2}$ and $\sigma_{2}$ for different $E$ with $h_{1}=4$
and $\sigma_{1}=70$. }
\end{figure}

\begin{itemize}
\item \textbf{Which Spatial Channel to Turn On?} Suppose $h_{1}=4$ and
$\sigma_{1}$ = 70, Figure \ref{fig: example-figure-1} illustrates
the number of spatial channels activated at different regions of the
state space $\left(h_{2},\sigma_{2}\right)$ for different $E$. It
can be observed that the region where both spatial channels are turned
on enlarges as the available energy $E$ increases. This is reasonable
because large $E$ means a good transmission opportunity which allows
more spatial channels to be activated. 
\end{itemize}

\section{STABILITY ANALYSIS}

In this section, we derive the sufficient condition for NCS stability
under the drift-minimizing MIMO linear precoding solution in Theorem
\ref{solution}, and analyze the MIMO plant state estimation performance.

\subsection{Sufficient Condition for NCS Stability}

Instability measures of the unstable dynamic plants provide key information
directly related to stabilizability and performance limitations. One
of the key measures that has been proposed in the literature for measuring
the instability of the unstable dynamic plants is defined as the product
of the modulus of the unstable eigenvalues of matrix $\mathbf{A}$
\cite{stein1989bode,qiu2010quantify}, i.e.,
\begin{align}
 & \mathcal{M}\left(\mathbf{A}\right)=\prod_{i=1}^{K}\mathrm{\mathrm{max\left(1,\left|\mu_{i}\left(\mathbf{A}\right)\right|\right)}}.\label{eq:insta-measure}
\end{align}
 In this section, we shall establish a sufficient condition for NCS
stability under the proposed MIMO precoding policy in terms of the
instability measure $\mathcal{M}\left(\mathbf{A}\right)$, the energy
arrival $\alpha\left(n\right)$ as well as the battery capacity $\theta$.
The sufficient condition for NCS stability is summarized in Theorem
2.

\begin{thm}
(\textsl{Sufficient Condition for NCS Stability)} \label{Thm suff condition}If
the following condition is satisfied:
\begin{align}
 & \mathbb{E}\left[\frac{1}{\alpha}\right]+\frac{1}{\theta}<\underset{\xi}{\mathrm{max}}\frac{1-\left(\varepsilon+K\mathrm{Pr}\left(\tilde{\pi}<\xi\right)\right)\mathcal{M}\left(\mathbf{A}\mathbf{A}^{T}\right)}{\delta^{2}K\tau\mathbb{E}\left[\left.\tilde{\pi}^{-1}\right|\tilde{\pi}\geq\xi\right]\mathcal{M}\left(\mathbf{A}\right){\normalcolor \mathcal{M}\left(\mathbf{A}\mathbf{A}^{T}\right)}},\label{eq:suff condition}
\end{align}
where $\tilde{\pi}=\frac{\pi}{\mathrm{Tr}\left(\boldsymbol{\Pi}_{K}^{-1}\right)}$
and $\pi$ is the unordered singular value of $\mathbf{H}$, $\delta=\sqrt{\frac{2}{\varepsilon}}\left(1+\left\Vert \mathbf{A}-\mathbf{B}\boldsymbol{\Psi}\mathbf{A}\right\Vert \Theta\right)\left\Vert \mathbf{B}\boldsymbol{\Psi}\right\Vert \left\Vert \mathbf{A}\right\Vert $
and $\Theta$ is a constant given in Lemma \ref{lemma-limiter-range},
then the dynamic MIMO plant is stable.\end{thm}

\begin{IEEEproof}
Please see Appendix E.
\end{IEEEproof}

The sufficient condition in Theorem 2 delivers some key design insights
into the dynamic MIMO plant system. Specifically, we have the following
system design insights.
\begin{itemize}
\item \textsl{Limiter Requirement:} The $\varepsilon$-saturation limiter
should be carefully designed such that the saturation probability
$\varepsilon$ obeys $\varepsilon<\mathcal{M}\left(\mathbf{A}\mathbf{A}^{T}\right)^{-1}-K\mathrm{Pr}\left(\tilde{\pi}<\xi^{\ast}\right)$,
where $\xi^{\ast}=\underset{\xi}{\mathrm{arg}\ \mathrm{max}}\ \frac{1-\left(\varepsilon+K\mathrm{Pr}\left(\tilde{\pi}<\xi\right)\right)\mathcal{M}\left(\mathbf{A}\mathbf{A}^{T}\right)}{\delta^{2}K\tau\mathbb{E}\left[\left.\tilde{\pi}^{-1}\right|\tilde{\pi}\geq\xi\right]\mathcal{M}\left(\mathbf{A}\right)\mathcal{M}\left(\mathbf{A}\mathbf{A}^{T}\right)}$.
This means that if the MIMO plant is very unstable (i.e., large $\mathcal{M}\left(\mathbf{A}\right)$),
the limiter should be designed with very small saturation probability
to guarantee stability. 
\item \textsl{Battery Capacity Requirement:} The battery capacity $\theta$
should obey $\theta>\frac{\delta^{2}K\tau\mathbb{E}\left[\left.\tilde{\pi}^{-1}\right|\tilde{\pi}\geq\xi\right]\mathcal{M}\left(\mathbf{A}\right)\mathcal{M}\left(\mathbf{A}\mathbf{A}^{T}\right)}{1-\left(\varepsilon+K\mathrm{Pr}\left(\tilde{\pi}<\xi^{\ast}\right)\right)\mathcal{M}\left(\mathbf{A}\mathbf{A}^{T}\right)}$.
This means the more unstable the MIMO plant is, the larger the capacity
of the energy storage devices is required. 
\item \textsl{Energy arrival requirement}: The sufficient condition also
implies that in order to achieve stability, there is a requirement
on the minimum average energy arrival rate, i.e., $\mathbb{E}\left[\alpha\right]\geq\left(\mathbb{E}\left[\frac{1}{\alpha}\right]\right)^{-1}>\left(\frac{1-\left(\varepsilon+K\mathrm{Pr}\left(\tilde{\pi}<\xi\right)\right)\mathcal{M}\left(\mathbf{A}\mathbf{A}^{T}\right)}{\delta^{2}K\tau\mathbb{E}\left[\left.\tilde{\pi}^{-1}\right|\tilde{\pi}\geq\xi\right]\mathcal{M}\left(\mathbf{A}\right)\mathcal{M}\left(\mathbf{A}\mathbf{A}^{T}\right)}-\frac{1}{\theta}\right)^{-1}$,
and larger average energy arrival rate (stronger energy harvesting
capability) is required for the more unstable MIMO plant. 
\end{itemize}

The sufficient condition (16) is not difficult to check because it
is expressed in terms of key system parameters from individual components
of the system. For example, In (16), $\delta$, $K$, $\mathcal{M}\left(\mathbf{A}\right)$
and $\mathcal{M}\left(\mathbf{A}\mathbf{A}^{T}\right)$ are the parameters
from the MIMO dynamic plant. $\mathbb{E}\left[\frac{1}{\alpha}\right]$
is the statistical parameters for renewable energy source, which can
be obtained from offline measurements. $\theta$ is the battery size
of the sensor. $\varepsilon$ is the sensor limiter saturation probability.
$\mathrm{Pr}\left(\tilde{\pi}<\xi\right)$ and $\mathbb{E}\left[\left.\tilde{\pi}^{-1}\right|\tilde{\pi}\geq\xi\right]$
are the statistical parameters for the MIMO fading channel, which
again can be obtained from offline measurements. 
\begin{remrk}
In \cite{stab-compare-1,stabi-compare-2}, the authors show that to
achieve stability there is a minimum data rate requirement in terms
of the instability measure (\ref{eq:insta-measure}). In our scenario,
if there is no sufficient energy storage, the sensor tends to be in
sleep mode and there is no data transmission between the sensor and
the controller. Hence, the available transmission energy at the sensor
will affect the communication data rate. This intuitively indicates
that to achieve stability there shall exist some requirement on the
available transmission energy at the sensor. This effect is shown
in the sufficient condition (\ref{eq:suff condition}), where there
is a requirement on the minimum average energy arrival rate in terms
of instability measure (\ref{eq:insta-measure}).
\end{remrk}

\subsection{State Estimation MSE Performance }

We are interested in analyzing the achievable state estimation MSE
using the proposed MIMO linear precoding policy. This is summarized
in the theorem below.

\begin{thm}
\textsl{(MSE of State Estimation Error)} \label{Thm MSE bound}If
the sufficient condition for NCS stability (\ref{eq:suff condition})
is satisfied, then the MSE satisfies: 
\begin{align}
 & \underset{N\rightarrow\infty}{\mathrm{limsup}}\frac{1}{N}\sum_{n=0}^{N-1}\mathbb{E}\left[\left\Vert \mathbf{x}\left(n\right)-\hat{\mathbf{x}}\left(n\right)\right\Vert ^{2}\right]\nonumber \\
 & \leq\frac{1}{\eta}\bigg(1+K\tau\frac{\delta^{2}}{\left\Vert \mathbf{B}\boldsymbol{\Psi}\right\Vert ^{2}}\mathbb{E}\left[\left.\tilde{\pi}^{-1}\right|\tilde{\pi}\geq\xi\right]\left(\mathbb{E}\left[\frac{1}{\alpha}\right]+\frac{1}{\theta}\right)\nonumber \\
 & \cdot\mathcal{M}\left(\mathbf{A}\right)\mathcal{M}\left(\mathbf{A}\mathbf{A}^{T}\right)\bigg)\mathrm{Tr}\left(\mathbf{W}\right)+\frac{\theta^{2}}{\eta},\label{eq:error bound}
\end{align}
where
\begin{align}
 & \eta=1-\varepsilon\mathcal{M}\left(\mathbf{A}\mathbf{A}^{T}\right)-K\mathrm{Pr}\left(\tilde{\pi}<\xi^{\ast}\right)\mathcal{M}\left(\mathbf{A}\mathbf{A}^{T}\right)\nonumber \\
 & -\left(\mathbb{E}\left[\frac{1}{\alpha}\right]+\frac{1}{\theta}\right)KT\delta^{2}\tau\mathbb{E}\left[\left.\tilde{\pi}^{-1}\right|\tilde{\pi}\geq\xi^{\ast}\right]\mathcal{M}\left(\mathbf{A}\right)\mathcal{M}\left(\mathbf{A}\mathbf{A}^{T}\right).
\end{align}
\end{thm}

\begin{IEEEproof}
Please see Appendix G.
\end{IEEEproof}

The state estimation performance bound (\ref{eq:error bound}) reveals
the fact that large average energy arrival rate (small $\mathbb{E}\left[\frac{1}{\alpha}\right]$)
and good MIMO channel quality (small $\mathbb{E}\left[\left.\tilde{\pi}^{-1}\right|\tilde{\pi}\geq\xi^{\ast}\right]$
and $\mathrm{Pr}\left(\tilde{\pi}<\xi^{\ast}\right)$) will result
in a better state estimation performance.

\section{Numerical Results}

In this section we compare the performance of the proposed MIMO linear
precoding scheme with the following baselines via numerical simulations.
\begin{itemize}
\item \textbf{Baseline 1} \emph{(MIMO Water-filling Precoding Maximizing
Capacity)} \cite{shi2007downlink,epi-lemma1}: The MIMO precoding
action is given by $\mathbf{F}\left(n\right)=\mathbf{U}\mathbf{\Upsilon}$,
where $\mathbf{\Upsilon}\in\mathbb{R}^{N_{t}\times K}$, $\left(\mathbf{\Upsilon}\right)_{ii}=\left[\gamma-\frac{1}{\left(\boldsymbol{\Pi}_{K}\right)_{ii}}\right]^{+}$
for all $i=1,\ldots,K$, the water level $\gamma$ is chosen such
that $\sum_{i=i}^{K}\left(\mathbf{\Upsilon}\right)_{ii}=\frac{E\left(n\right)}{M^{2}\tau}$,
and the other elements of $\mathbf{\Upsilon}$ are zero. This corresponds
to the physical layer capacity maximizing solution. 
\item \textbf{Baseline 2} \emph{(Periodic MIMO Water-filling Precoding)}:
The sensor is periodically activated for transmission with a fixed
period of $T$ . The sensor adopts the MIMO water-filling precoding
in baseline 1 when it transmits.
\item \textbf{Baseline 3} \emph{(MIMO Water-filling Precoding Minimizing
MSE) }\cite{mmse-precoder}: The MSE minimizing precoding solution
is given by $\mathbf{F}\left(n\right)=\mathbf{U}\mathbf{\Upsilon}$,
where $\left(\mathbf{\Upsilon}\right)_{ii}=\left[\frac{\gamma}{\sqrt{\left(\boldsymbol{\Pi}_{K}\right)_{ii}}}-\frac{1}{\left(\boldsymbol{\Pi}_{K}\right)_{ii}}\right]^{+}$
for all $i=1,\ldots,K$, and $\gamma$ is chosen such that $\sum_{i=i}^{K}\left(\mathbf{\Upsilon}\right)_{ii}=\frac{E\left(n\right)}{M^{2}\tau}$,
and the other elements of $\mathbf{\Upsilon}$ are zero. 
\item \textbf{Baseline 4 }\emph{(MIMO Water-filling Precoding Maximizing
Capacity with Constant Power Supply)}: The MIMO precoding is in the
same form as Baseline 1 except that the MIMO precoding is supplied
with constant power $\mathbb{E}\left[\alpha\right]$, i.e., $\sum_{i=i}^{K}\left(\mathbf{\Upsilon}\right)_{ii}=\frac{\mathbb{E}\left[\alpha\right]}{M^{2}\tau}$,
where the constant $\mathbb{E}\left[\alpha\right]$ is the average
energy arrival.
\item \textbf{Baseline 5} \emph{(MIMO Water-filling Precoding Minimizing
MSE with Constant Power Supply)}: The MIMO precoding is in the same
form as Baseline 3 except that the MIMO precoding is supplied with
constant power $\mathbb{E}\left[\alpha\right]$, i.e., $\sum_{i=i}^{K}\left(\mathbf{\Upsilon}\right)_{ii}=\frac{\mathbb{E}\left[\alpha\right]}{M^{2}\tau}$,
where the constant $\mathbb{E}\left[\alpha\right]$ is the average
energy arrival.
\end{itemize}

We consider a MIMO with parameters: $\mathbf{A}=\left(\begin{array}{cc}
1.3 & 0.1\\
-0.2 & 1.2
\end{array}\right)$, $\mathbf{\mathbf{B}}=\mathrm{diag(1,1)}$, $\mathbf{\mathbf{W}}=\mathrm{diag(1,2)}$,
$\mathbf{\mathbf{Q}}=\mathrm{diag(1,1)}$, $\mathbf{\mathbf{\Psi}}=\mathrm{diag(0.25,0.25)}$,
$\varepsilon=0.05$, $M=1$, $N_{s}=3$, $N_{c}=2$, $T=3\tau$ and
$\tau=0.01s$. The harvestable energy process $\alpha\left(n\right)$
is assumed to be Poisson distributed. The MIMO channel $\mathbf{H}\left(n\right)$
is a $2\times3$ matrix with each element being i.i.d. complex Gaussian
distributed with zero mean and unit variance. The MIMO plant dynamics
$\mathbf{A}$ has unstable eigenvalues with instability measure $\mathcal{M}\left(\mathbf{A}\right)=1.58$.
The simulation is Monte Carlo simulation and for a given pair of average
energy arrival $\mathbb{E}\left[\alpha\right]$ and battery capacity
$\theta$, we simulate 5000 sample paths of plant state evolution
and estimation, and each sample path contains 300 time slots. The
system parameter configurations, namely the harvestable energy process
$\alpha\left(n\right)$, battery capacity $\theta$, MIMO channel
$\mathbf{H}$, limiter saturation probability $\varepsilon$, and
limiter parameter $\delta$ of Baseline 1-3 are the same as the proposed
scheme, and the system parameter configurations satisfy the stability
condition (\ref{eq:suff condition}) in Theorem 2. The Baseline 1-3
and the proposed scheme also satisfy the peak power restriction and
the energy harvesting constraints.

\subsection{State Estimation MSE Versus Battery Capacity $\theta$}

\begin{figure}[tbh]
\begin{centering}
\includegraphics[clip,width=1\columnwidth]{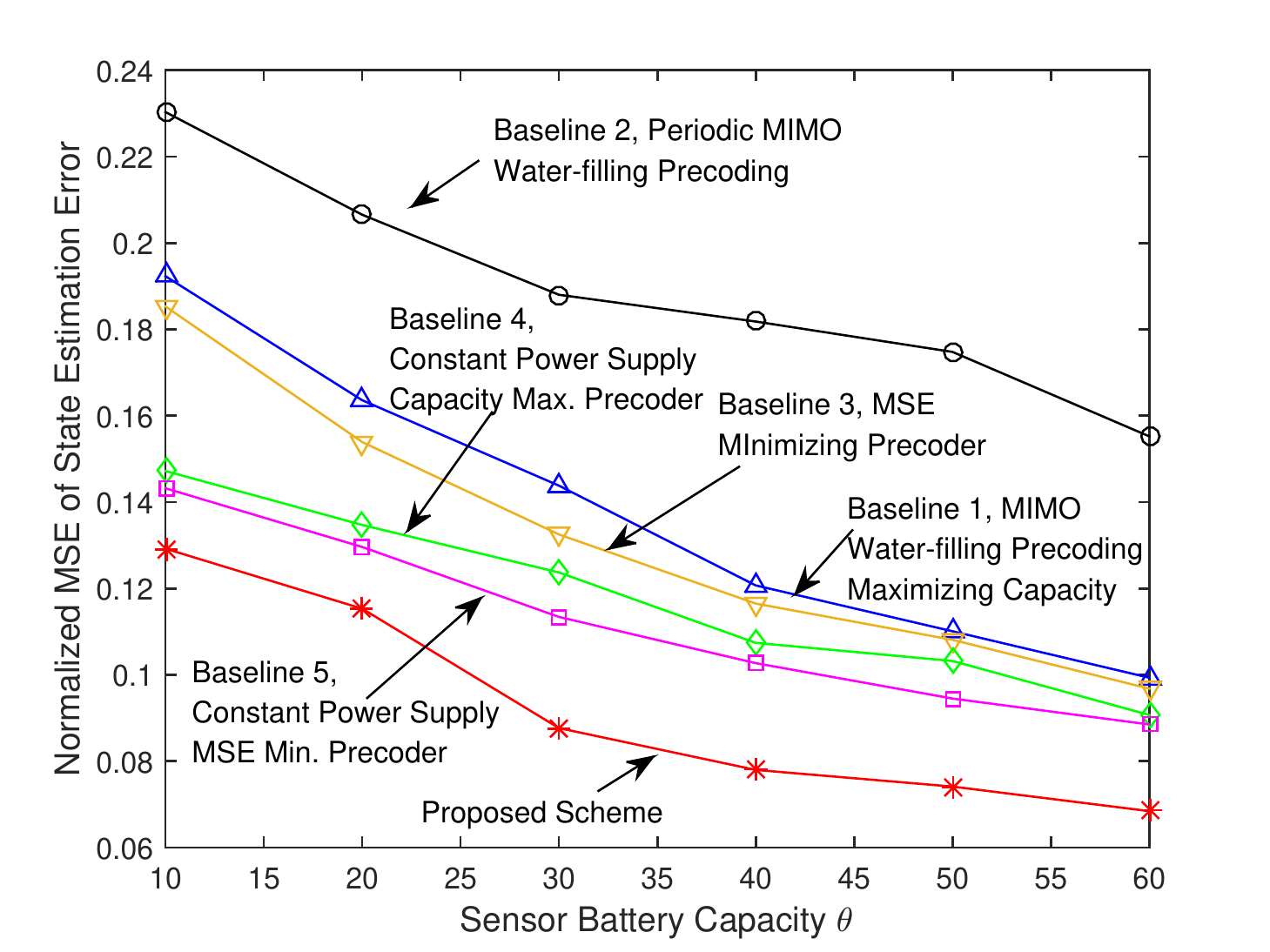}
\par\end{centering}

\caption{\label{fig: vs-theta}Normalized state estimation error versus battery
capacity $\theta$ under $\mathbb{E}\left[\alpha\right]={\normalcolor 40J}$.}
\end{figure}

Figure \ref{fig: vs-theta} illustrates the normalized MSE of the
state estimation error versus the sensor battery capacity $\theta$
under the average energy arrival $\mathbb{E}\left[\alpha\right]=40{\color{blue}{\normalcolor J}}$.
Note that Baseline 1 is the MIMO precoding that adapts to the CSI
and energy state only and maximizes the physical layer capacity. Baseline
2 covers a lot of existing NCS literature where a static transmission
scheme with the sensor periodically being active is adopted \cite{li2013optimal}.
Baseline 3 is the MIMO precoding that maximizes the receiver MSE.
Baseline 4 and 5 are supplied with constant power $\mathbb{E}\left[\alpha\right]$
to compare the effect of energy harvesting constraints. 

It can be observed that there is a significant performance gain of
the proposed scheme compared with all the baselines. This is because
the proposed scheme is plant state and CSI aware and is tailored for
NCS applications. Baseline 1 is not tailored for NCS applications
in the sense that Baseline 1 merely optimizes the physical layer throughput
in wireless communications. Baseline 2 has the worst performance because
it neither fully exploits the CSI nor adapts to the plant state. Baseline
3 has better MSE performance compared with the capacity maximizing
solution (Baseline 1). The proposed scheme outperforms Baseline 3
because Baseline 3 still fails to exploit the transmission urgency
of the MIMO plant state. Baseline 4 and 5, which is supplied with
constant power $\mathbb{E}\left[\alpha\right]$, have better MSE performance
compared with the Baselines have energy harvesting constraints, but
Baseline 4 and 5 still fail to exploit state transmission urgency
of the MIMO dynamic plant. 

\subsection{State Estimation MSE Versus Average Energy Arrival $\mathbb{E}\left[\alpha\right]$}

\begin{figure}[tbh]
\begin{centering}
\includegraphics[clip,width=1\columnwidth]{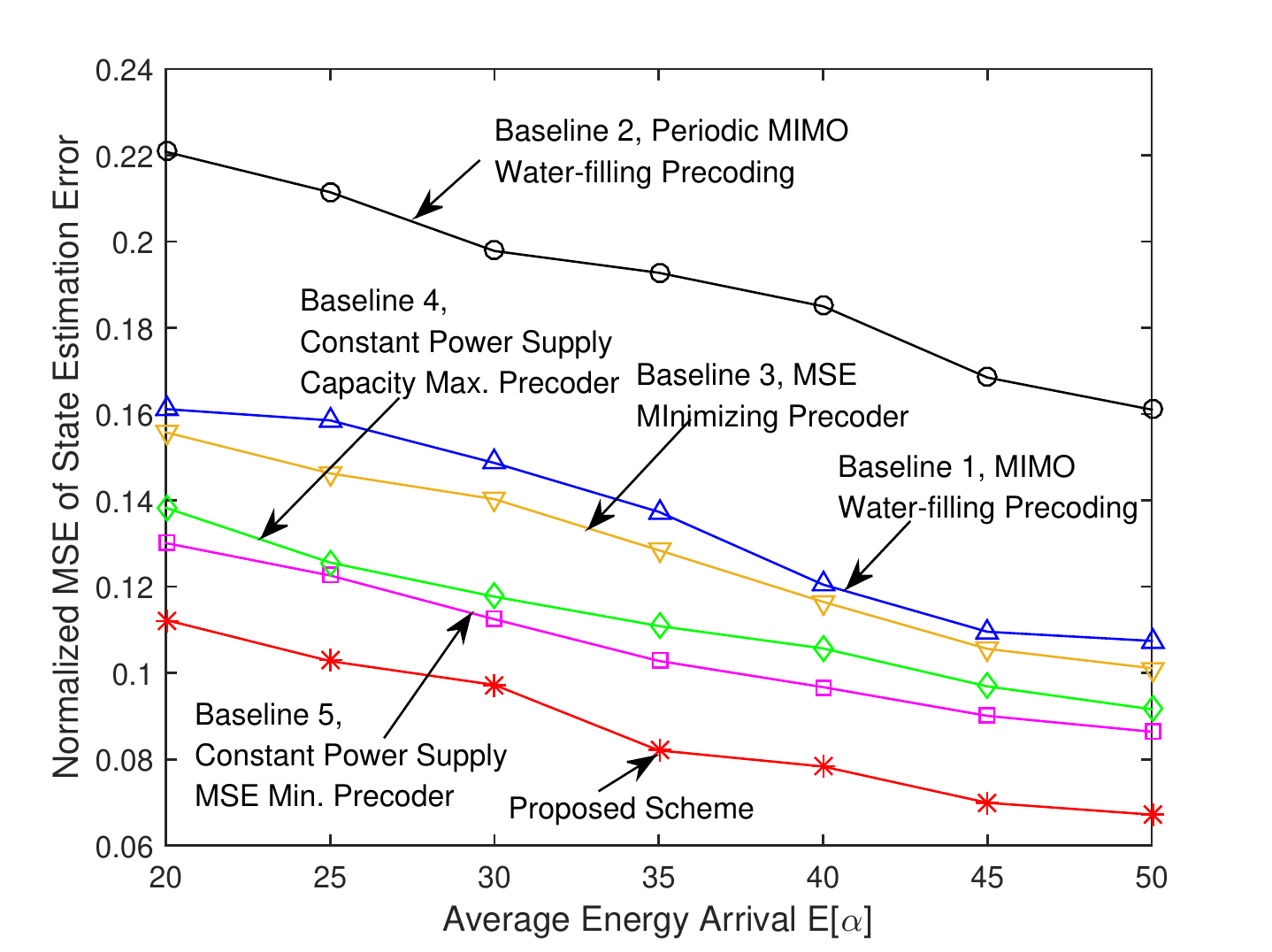}
\par\end{centering}

\caption{\label{fig: vs-alpha}Normalized state estimation error versus average
energy arrival $\mathbb{E}\left[\alpha\right]=40J$ under $\theta=80J$.}
\end{figure}

Figure \ref{fig: vs-alpha} illustrates the normalized MSE of the
state estimation error versus the average energy arrival $\mathbb{E}\left[\alpha\right]$
with sensor battery capacity $\theta=80J$. It can be observed that
the state estimation MSE decreases as the average energy arrival increases.
And the proposed scheme has a large performance gain compared with
all the baselines.\textcolor{blue}{{} }

\section{Conclusion}

In this paper, we consider an NCS with an energy harvesting sensor
that delivers the MIMO dynamic plant state information to a controller
over a MIMO fading channel. We derived a low-complexity closed-form
dynamic MIMO precoder at the sensor via Lyapunov optimization to stabilize
the unstable MIMO dynamic plant. The solution has an event-driven
structure and an eigenvalue water-filling structure. We also analyzed
the sufficient condition for NCS stability under the proposed MIMO
precoder solution. Compared with the various existing MIMO linear
precoding schemes, the proposed MIMO precoder solution has substantial
state estimation performance gain.

\appendix

\subsection{\label{sub:Proof-of-mse-bound}Proof of Lemma \ref{lemma-mse-bound}}

Denote the history of the realizations of variable $a$ up to time
slot $n$ as $a_{0}^{n}\triangleq\cup_{i=1}^{n}\left\{ a\left(i\right)\right\} .$
At time slot $n$, the knowledge available to the controller is given
by the information set $I_{c}\left(n\right)=\big\{\mathbf{u}_{0}^{n-1},\mathbf{\tilde{F}}_{0}^{n},\mathbf{y}_{0}^{n},\gamma_{0}^{n}\big\},\ \forall n\in\mathbb{N}^{+}$,
with $I_{c}(0)=\big\{\tilde{\mathbf{F}}(0),\mathbf{y}(0),\gamma\left(0\right)\big\}.$
Let the information set $\widetilde{I}_{c}\left(n\right)=\big\{\mathbf{\tilde{F}}_{0}^{n},\mathbf{\tilde{y}}_{0}^{n},\gamma_{0}^{n}\big\}$,
where $\tilde{\mathbf{y}}(n)=\mathbf{y}\left(n\right)\gamma\left(n\right)$.
We define a covariance matrix $\boldsymbol{\Sigma}\left(n\right)=\mathbb{E}\left[\left.\left(\mathbf{x}\left(n\right)-\widetilde{\hat{\mathbf{x}}}\left(n\right)\right)\left(\mathbf{x}\left(n\right)-\widetilde{\hat{\mathbf{x}}}\left(n\right)\right)^{T}\right|\widetilde{I}_{c}\left(n-1\right)\right]$,
where $\widetilde{\hat{\mathbf{x}}}\left(n\right)=\mathbb{E}\left[\left.\mathbf{x}\left(n\right)\right|\widetilde{I}_{c}\left(n-1\right)\right]$.
Define $\boldsymbol{\Sigma}^{+}\left(n\right)=\mathbb{E}\left[\left.\left(\mathbf{x}\left(n\right)-\widetilde{\hat{\mathbf{x}}}^{+}\left(n\right)\right)\left(\mathbf{x}\left(n\right)-\widetilde{\hat{\mathbf{x}}}^{+}\left(n\right)\right)^{T}\right|\widetilde{I}_{c}\left(n\right)\right]$,
where $\widetilde{\hat{\mathbf{x}}}^{+}\left(n\right)=\mathbb{E}\left[\left.\mathbf{x}\left(n\right)\right|\widetilde{I}_{c}\left(n\right)\right]$.
Consider a virtual scenario setup where the virtual noisy plant state
measurements $\mathbf{y}\left(n\right)=\mathbf{\mathbf{\mathbf{\tilde{F}}}}\left(n\right)\mathbf{x}\left(n\right)+\mathbf{z}\left(n\right)$
are sent across a packet-dropping channel. The packet dropping channel
is modeled by $\tilde{\mathbf{y}}(n)=\gamma\left(n\right)\mathbf{y}\left(n\right)$
\cite{optimal-kf-power-allo}. This virtual scenario setup coincides
with the scenario studied in \cite{sinopoli2004kalmanintermit} and
by the same argument in \cite{sinopoli2004kalmanintermit}, the conditional
probability density function $p\left(\left.\mathbf{z}\left(n\right)\right|\gamma\left(n\right)\right)$
of $\mathbf{z}\left(n\right)$ is given by: if $\gamma\left(n\right)=1$,
$p\left(\left.\mathbf{z}\left(n\right)\right|\gamma\left(n\right)\right)=\mathcal{CN}\left(0,\mathbf{I}\right);$
otherwise, $p\left(\left.\mathbf{z}\left(n\right)\right|\gamma\left(n\right)\right)=\mathcal{CN}\left(0,\sigma\mathbf{I}\right).$
According to the augmented complex Kalman filter algorithm in \cite{augmented},
$\boldsymbol{\Sigma}\left(n+1\right)$ and $\boldsymbol{\Sigma}^{+}\left(n+1\right)$can
be obtained recursively as follows \cite{sinopoli2004kalmanintermit,augmented}:
\begin{align}
\boldsymbol{\Sigma}\left(n+1\right) & =\mathbf{A}\boldsymbol{\Sigma}^{+}\left(n\right)\mathbf{A}^{T}+\mathbf{W};\label{eq:1}\\
\boldsymbol{\Sigma}^{+}\left(n+1\right) & =\boldsymbol{\Sigma}\left(n+1\right)-\boldsymbol{\Sigma}\left(n+1\right)\left(\mathbf{\mathbf{\mathbf{\tilde{F}}}}^{a}\left(n\right)\right)^{H}\nonumber \\
 & \bigg(\mathbf{\mathbf{\mathbf{\tilde{F}}}}^{a}\left(n\right)\boldsymbol{\Sigma}\left(n\right)\big(\mathbf{\mathbf{\mathbf{\tilde{F}}}}^{a}\left(n\right)\big)^{H}+\gamma\left(n+1\right)\mathbf{I}\nonumber \\
 & +\left(1-\gamma\left(n+1\right)\right)\sigma\mathbf{I}\bigg)^{-1}\mathbf{\mathbf{\mathbf{\tilde{F}}}}^{a}\left(n\right)\boldsymbol{\Sigma}\left(n+1\right)\label{eq:2}
\end{align}
Combining equation (\ref{eq:1}) and (\ref{eq:2}), and taking the
limit as $\sigma\rightarrow\infty,$ it follows that the dynamics
$\mathbf{\boldsymbol{\Sigma}}\left(n\right)$ is given by equation
(\ref{dyna-sigma}).

Note that given $I_{c}\left(n-1\right)$, $\widetilde{I}_{c}\left(n-1\right)$
is independent of $\mathbf{x}\left(n\right)$. Hence $\mathbf{x}\left(n\right)$,
$I_{c}\left(n-1\right)$ and $\widetilde{I}_{c}\left(n-1\right)$
form a Markov chain $\mathbf{x}\left(n\right)\rightarrow I_{c}\left(n-1\right)\rightarrow\widetilde{I}_{c}\left(n-1\right)$
\textcolor{blue}{}\footnote{Random variables $X$, $Y$, $Z$ are said to form a Markov chain
in that order (denoted by $X\rightarrow Y\rightarrow Z$) if the conditional
distribution of $Z$ depends only on $Y$ and is conditionally independent
of $X$. $X\rightarrow Y\rightarrow Z$ if and only if $X$ and $Z$
are conditionally independent given $Y$ \cite{epi-lemma1}.}. Hence
\begin{align}
 & \mathbb{E}\left[\left.\mathbf{x}\left(n\right)\right|\widetilde{I}_{c}\left(n-1\right)\right]=\mathbb{E}\left[\left.\mathbb{E}\left[\left.\mathbf{x}\left(n\right)\right|I_{c}\left(n-1\right)\right]\right|\widetilde{I}_{c}\left(n-1\right)\right].\label{eq:E1}
\end{align}
By the law of total covariance \cite{O.Rioul}, it follows:
\begin{align}
 & \mathrm{Cov}\left[\mathbb{E}\left[\left.\mathbf{x}\left(n\right)\right|I_{c}\left(n-1\right)\right]\right]\nonumber \\
 & =\mathrm{Cov}\left[\left.\mathbb{E}\left[\left.\mathbf{x}\left(n\right)\right|I_{c}\left(n-1\right)\right]\right|\widetilde{I}_{c}\left(n-1\right)\right]\nonumber \\
 & +\mathrm{Cov}\left[\mathbb{E}\left[\left.\mathbb{E}\left[\left.\mathbf{x}\left(n\right)\right|I_{c}\left(n-1\right)\right]\right|\widetilde{I}_{c}\left(n-1\right)\right]\right].\label{eq:E3-1}
\end{align}
Substituting (21) into (22), it follows:
\begin{align}
 & \mathrm{Cov}\left[\mathbb{E}\left[\left.\mathbf{x}\left(n\right)\right|I_{c}\left(n-1\right)\right]\right]\nonumber \\
 & =\mathrm{Cov}\left[\left.\mathbb{E}\left[\left.\mathbf{x}\left(n\right)\right|I_{c}\left(n-1\right)\right]\right|\widetilde{I}_{c}\left(n-1\right)\right]\nonumber \\
 & +\mathrm{Cov}\left[\mathbb{E}\left[\left.\mathbf{x}\left(n\right)\right|\widetilde{I}_{c}\left(n-1\right)\right]\right].\label{eq:E3}
\end{align}
Furthermore, 
\begin{align}
 & \mathrm{Cov}\left[\mathbb{E}\left[\left.\mathbf{x}\left(n\right)\right|I_{c}\left(n-1\right)\right]\right]-\mathrm{Cov}\left[\mathbb{E}\left[\left.\mathbf{x}\left(n\right)\right|\widetilde{I}_{c}\left(n-1\right)\right]\right]\nonumber \\
 & =\mathrm{Cov}\left[\left.\mathbf{x}\left(n\right)\right|\widetilde{I}_{c}\left(n-1\right)\right]-\mathrm{Cov}\left[\left.\mathbf{x}\left(n\right)\right|I_{c}\left(n-1\right)\right].\label{eq:E6}
\end{align}
Substituting (\ref{eq:E6}) into (\ref{eq:E3}), it follows:
\begin{align}
 & \mathrm{Cov}\left[\left.\mathbf{x}\left(n\right)\right|I_{c}\left(n-1\right)\right]=\mathrm{Cov}\left[\left.\mathbf{x}\left(n\right)\right|\widetilde{I}_{c}\left(n-1\right)\right]\nonumber \\
 & -\mathrm{Cov}\left[\left.\mathbb{E}\left[\left.\mathbf{x}\left(n\right)\right|I_{c}\left(n-1\right)\right]\right|\widetilde{I}_{c}\left(n-1\right)\right].\label{eq:E7}
\end{align}
Since covariance matrices are positive semidefinite, it follows: 
\begin{align}
\mathrm{Cov}\left[\left.\mathbf{x}\left(n\right)\right|I_{c}\left(n-1\right)\right] & \leq\boldsymbol{\Sigma}\left(n\right).\label{eq:E8}
\end{align}
We construct another Markov chain $\mathbf{x}\left(n\right)\rightarrow I_{c}\left(n\right)\rightarrow I_{c}\left(n-1\right).$
Note that $I_{c}\left(n-1\right)\subseteq I_{c}\left(n\right).$ Therefore,
given $I_{c}\left(n\right)$, $I_{c}\left(n-1\right)$ is constant
and is independent of $\mathbf{x}\left(n\right)$. Therefore, by definition
in Section 2.8 of \cite{epi-lemma1}, $\mathbf{x}\left(n\right)$,
$I_{c}\left(n\right)$ and $I_{c}\left(n-1\right)$ form a Markov
chain $\mathbf{x}\left(n\right)\rightarrow I_{c}\left(n\right)\rightarrow I_{c}\left(n-1\right)$.
Applying Proposition 5 in \cite{O.Rioul}, it follows:
\begin{align}
 & \mathbb{E}\left[\left.\left(\mathbf{x}\left(n\right)-\hat{\mathbf{x}}\left(n\right)\right)\left(\mathbf{x}\left(n\right)-\hat{\mathbf{x}}\left(n\right)\right)^{T}\right|I_{c}\left(n\right)\right]\nonumber \\
 & =\mathrm{Var}\left[\left.\mathbf{x}\left(n\right)\right|I_{c}\left(n\right)\right]\leq\mathrm{Var}\left[\left.\mathbf{x}\left(n\right)\right|I_{c}\left(n-1\right)\right]\nonumber \\
 & =\mathbb{E}\left[\left.\left(\mathbf{x}\left(n\right)-\hat{\mathbf{x}}^{-}\left(n\right)\right)\left(\mathbf{x}\left(n\right)-\hat{\mathbf{x}}^{-}\left(n\right)\right)^{T}\right|I_{c}\left(n-1\right)\right],\label{eq:E9}
\end{align}
where $\hat{\mathbf{x}}^{-}\left(n\right)=\mathbb{E}\left[\left.\mathbf{x}\left(n\right)\right|I_{c}\left(n-1\right)\right]$. 

Combining (\ref{eq:E8}) and (\ref{eq:E9}), and taking the trace
and expectation w.r.t. $I_{c}\left(n\right)$, it follows the inequality
(\ref{eq:mse-bound}). Therefore, Lemma \ref{lemma-mse-bound} is
proved.

\subsection{\label{sub:Proof-of-connection}Proof of Lemma \ref{lemma-connnection}}

Note that $\mathbf{x}\left(n+1\right)=\mathbf{A}\mathbf{x}\left(n\right)-\mathbf{B}\boldsymbol{\Psi}\mathbf{A}\widetilde{\hat{\mathbf{x}}}\left(n\right)+\mathbf{w}\left(n\right)$,
it follows:
\begin{align}
\left\Vert \mathbf{x}\left(n+1\right)\right\Vert ^{2} & \mathbb{\leq}\left\Vert \mathbf{A}-\mathbf{B}\boldsymbol{\Psi}\mathbf{A}\right\Vert ^{2}\mathbb{E}\left[\left\Vert \mathbf{x}\left(n\right)\right\Vert ^{2}\right]\nonumber \\
 & +\mu_{max}\left(\mathbf{A}\mathbf{A}^{T}\right)\mathbb{E}\left[\left\Vert \mathbf{x}\left(n\right)-\widetilde{\hat{\mathbf{x}}}\left(n\right)\right\Vert ^{2}\right]+\mathrm{Tr}\left(\mathbf{W}\right).
\end{align}
Since $\mathbf{A}-\mathbf{B}\boldsymbol{\Psi}\mathbf{A}$ is a Hurwitz
matrix, then $\left\Vert \mathbf{A}-\mathbf{B}\boldsymbol{\Psi}\mathbf{A}\right\Vert <1$.
If $\left\Vert \mathbf{A}-\mathbf{B}\boldsymbol{\Psi}\mathbf{A}\right\Vert =0$,
then $\left\Vert \mathbf{x}\left(n+1\right)\right\Vert ^{2}<\mu_{max}\left(\mathbf{A}\mathbf{A}^{T}\right)\mathbb{E}\left[\left\Vert \mathbf{x}\left(n\right)-\widetilde{\hat{\mathbf{x}}}\left(n\right)\right\Vert ^{2}\right]+\mathrm{Tr}\left(\mathbf{W}\right).$
Combining with Lemma \ref{lemma-mse-bound} and note that $\mathbb{E}\left[\left\Vert \mathbf{x}\left(n\right)-\widetilde{\hat{\mathbf{x}}}\left(n\right)\right\Vert ^{2}\right]=\mathbb{E}\left[\mathrm{Tr}\left(\boldsymbol{\Sigma}\left(n\right)\right)\right]$,
it follows: 
\begin{align}
 & \underset{N\rightarrow\infty}{\mathrm{limsup}}\frac{1}{N}\sum_{n=1}^{N}\mathbb{E}\left[\left\Vert \mathbf{x}\left(n+1\right)\right\Vert ^{2}\right]\nonumber \\
 & <\frac{\mu_{max}\left(\mathbf{A}\mathbf{A}^{T}\right)}{N}\underset{N\rightarrow\infty}{\mathrm{limsup}}\sum_{n=1}^{N}\mathbb{E}\left[\mathrm{Tr}\left(\boldsymbol{\Sigma}\left(n\right)\right)\right]+\mathrm{Tr}\left(\mathbf{W}\right)<\infty.
\end{align}
 Similarly, if $\left\Vert \mathbf{A}-\mathbf{B}\boldsymbol{\Psi}\mathbf{A}\right\Vert \neq0$,
then 
\begin{align}
 & \underset{N\rightarrow\infty}{\mathrm{limsup}}\frac{1}{N}\sum_{n=1}^{N}\mathbb{E}\left[\left\Vert \mathbf{x}\left(n+1\right)\right\Vert ^{2}\right]\nonumber \\
 & <\frac{\mu_{max}\left(\mathbf{A}\mathbf{A}^{T}\right)}{N\left(1-\left\Vert \mathbf{A}-\mathbf{B}\boldsymbol{\Psi}\mathbf{A}\right\Vert \right)}\underset{N\rightarrow\infty}{\mathrm{limsup}}\sum_{n=1}^{N}\mathbb{E}\left[\mathrm{Tr}\left(\boldsymbol{\Sigma}\left(n\right)\right)\right]\nonumber \\
 & +\frac{\mathrm{Tr}\left(\mathbf{W}\right)}{1-\left\Vert \mathbf{A}-\mathbf{B}\boldsymbol{\Psi}\mathbf{A}\right\Vert }<\infty.
\end{align}
Therefore, Lemma \ref{lemma-connnection} is proved.

\subsection{\label{sub:Proof-of-limiter-range}Proof of Lemma \ref{lemma-limiter-range}}

By the standard Lyapunov stability theory for discrete-time linear
systems, there exist positive definite symmetric matrices $\mathbf{Q}$
and $\mathbf{T}$ such that $\left(\mathbf{A}-\mathbf{B}\mathbf{\Psi}\mathbf{A}\right)^{T}\mathbf{Q}\left(\mathbf{A}-\mathbf{B}\mathbf{\Psi}\mathbf{A}\right)-\mathbf{Q}=-\mathbf{T}$.
It follows \cite{limiterproof}:
\begin{align*}
 & \mathbf{x}^{T}\left(n+1\right)\mathbf{S}\mathbf{x}\left(n+1\right)-\mathbf{x}^{T}\left(n\right)\mathbf{S}\mathbf{x}\left(n\right)\\
\leq & -\mu_{min}\left(\mathbf{T}\right)\left\Vert \mathbf{x}\left(n\right)\right\Vert ^{2}+2\left\Vert \mathbf{x}\left(n\right)\right\Vert \left\Vert \left(\mathbf{A}-\mathbf{B}\mathbf{\Psi}\mathbf{A}\right)^{T}\mathbf{Q}\right\Vert \\
 & \cdot\left(\left\Vert \mathbf{B}\mathbf{\Psi}\mathbf{A}\right\Vert \left\Vert \mathbf{x}\left(n\right)-\widetilde{\hat{\mathbf{x}}}\left(n\right)\right\Vert +\left\Vert \mathbf{w}\left(n\right)\right\Vert \right)\\
 & +\left\Vert \mathbf{Q}\right\Vert \left(\left\Vert \mathbf{B}\mathbf{\Psi}\mathbf{A}\right\Vert \left\Vert \mathbf{x}\left(n\right)-\widetilde{\hat{\mathbf{x}}}\left(n\right)\right\Vert +\left\Vert \mathbf{w}\left(n\right)\right\Vert \right)^{2}.
\end{align*}
Define
\begin{align*}
 & \Theta=\frac{1}{\mu_{min}\left(\mathbf{T}\right)}\bigg(\left\Vert \left(\mathbf{A}-\mathbf{B}\mathbf{\Psi}\mathbf{A}\right)^{T}\mathbf{Q}\right\Vert \\
 & +\left(\left\Vert \left(\mathbf{A}-\mathbf{B}\mathbf{\Psi\mathbf{A}}\right)^{T}\mathbf{Q}\right\Vert ^{2}+\mu_{min}\left(\mathbf{T}\right)\left\Vert \mathbf{Q}\right\Vert \right)^{\nicefrac{1}{2}}\bigg)\bigg),
\end{align*}
 it follows:
\begin{itemize}
\item If $\left\Vert \mathbf{x}\left(n\right)\right\Vert <\Theta\left(\left\Vert \mathbf{B}\mathbf{\Psi}\mathbf{A}\right\Vert \left\Vert \mathbf{x}\left(n\right)-\widetilde{\hat{\mathbf{x}}}\left(n\right)\right\Vert +\left\Vert \mathbf{w}\left(n\right)\right\Vert \right),$
although the one step drift $\left(\mathbf{x}^{T}\left(n+1\right)\mathbf{Q}\mathbf{x}\left(n+1\right)-\mathbf{x}^{T}\left(n\right)\mathbf{Q}\mathbf{x}\left(n\right)\right)$
is positive, we have
\begin{align}
 & \left\Vert \mathbf{x}\left(n+1\right)\right\Vert \leq\left(1+\left\Vert \mathbf{A}-\mathbf{B}\mathbf{\Psi}\mathbf{A}\right\Vert \Theta\right)\nonumber \\
 & (\left\Vert \mathbf{B}\mathbf{\Psi}\mathbf{A}\right\Vert \left\Vert \mathbf{x}\left(n\right)-\widetilde{\hat{\mathbf{x}}}\left(n\right)\right\Vert +\left\Vert \mathbf{w}\left(n\right)\right\Vert ).\label{eq:x_n+1-bound}
\end{align}

\item If $\left\Vert \mathbf{x}\left(n\right)\right\Vert >\Theta\left(\left\Vert \mathbf{B}\mathbf{\Psi}\mathbf{A}\right\Vert \left\Vert \mathbf{x}\left(n\right)-\widetilde{\hat{\mathbf{x}}}\left(n\right)\right\Vert +\left\Vert \mathbf{w}\left(n\right)\right\Vert \right),$
the one step drift $\mathbf{x}^{T}\left(n+1\right)\mathbf{Q}\mathbf{x}\left(n+1\right)-\mathbf{x}^{T}\left(n\right)\mathbf{Q}\mathbf{x}\left(n\right)$
is negative, hence $\mathbf{x}\left(n+1\right)$ will stay in the
ball:
\begin{align*}
\bigg\{\mathbf{x}\left(n+1\right):\left\Vert \mathbf{x}\left(n+1\right)\right\Vert \leq\left(1+\left\Vert \mathbf{A}-\mathbf{B}\mathbf{\Psi}\mathbf{A}\right\Vert \Theta\right)\\
\left(\left\Vert \mathbf{B}\mathbf{\Psi}\mathbf{A}\right\Vert \left\Vert \mathbf{x}\left(n\right)-\widetilde{\hat{\mathbf{x}}}\left(n\right)\right\Vert +\left\Vert \mathbf{w}\left(n\right)\right\Vert \right)\bigg\}.
\end{align*}
Hence we have $\left\Vert \mathbf{x}\left(n\right)\right\Vert \leq\left(1+\left\Vert \mathbf{A}-\mathbf{B}\mathbf{\Psi}\right\Vert \Theta\right)\left(\left\Vert \mathbf{B}\mathbf{\Psi}\mathbf{A}\right\Vert \left\Vert \mathbf{x}\left(n\right)-\widetilde{\hat{\mathbf{x}}}\left(n\right)\right\Vert +\left\Vert \mathbf{w}\left(n\right)\right\Vert \right)$
holds for any time slot $n$. 
\end{itemize}
Note that $\mathrm{Pr}\left[\left.\left\Vert \mathbf{x}\left(n\right)\right\Vert ^{2}>L^{2}\left(n\right)\right|I_{c}\left(n-1\right)\right]=\mathbb{E}\left[\left.\mathbf{1}\left\{ \left\Vert \mathbf{x}\left(n\right)\right\Vert ^{2}>L^{2}\left(n\right)\right\} \right|I_{c}\left(n-1\right)\right]$.
Suppose if $\mathbb{E}\left[\left.\mathbf{1}\left\{ \left\Vert \mathbf{x}\left(n\right)\right\Vert ^{2}>L^{2}\left(n\right)\right\} \right|I_{c}\left(n-1\right)\right]\leq\varepsilon$
holds, then taking expectation on both sides w.r.t. the randomness
of $I_{c}\left(n-1\right)$ will yield $\mathbb{E}\left[\mathbf{1}\left\{ \left\Vert \mathbf{x}\left(n\right)\right\Vert ^{2}>L^{2}\left(n\right)\right\} \right]=\mathrm{Pr}\left[\left\Vert \mathbf{x}\left(n\right)\right\Vert ^{2}>L^{2}\left(n\right)\right]\leq\varepsilon$.
Therefore, in order to guarantee the saturation probability of the
limiter is at most $\varepsilon$, it is sufficient to design the
limiter range $L\left(n\right)$ such that $\mathbb{E}\left[\left.\mathbf{1}\left\{ \left\Vert \mathbf{x}\left(n\right)\right\Vert ^{2}>L^{2}\left(n\right)\right\} \right|I_{c}\left(n-1\right)\right]\leq\varepsilon$
holds for all $n$.

By the Markov's inequality, it follows:
\begin{align}
\mathrm{Pr}\left[\left.\left\Vert \mathbf{x}\left(n\right)\right\Vert ^{2}>L^{2}\left(n\right)\right|I_{c}\left(n-1\right)\right] & \leq\frac{\mathbb{E}\left[\left.\left\Vert \mathbf{x}\left(n\right)\right\Vert ^{2}\right|I_{c}\left(n-1\right)\right]}{L^{2}\left(n\right)}.\label{eq:Markov-ineq}
\end{align}
Substituting (\ref{eq:x_n+1-bound}) into (\ref{eq:Markov-ineq})
and applying Lemma (\ref{lemma-mse-bound}), it follows: 
\begin{align}
 & \frac{\mathbb{E}\left[\left.\left\Vert \mathbf{x}\left(n\right)\right\Vert ^{2}\right|I_{c}\left(n-1\right)\right]}{L^{2}\left(n\right)}\nonumber \\
 & \leq\frac{1}{L^{2}\left(n\right)}\left(1+\left\Vert \mathbf{A}-\mathbf{B}\boldsymbol{\Psi}\mathbf{A}\right\Vert \Theta\right)^{2}\nonumber \\
 & \cdot\left(\left\Vert \mathbf{B}\boldsymbol{\Psi}\mathbf{A}\right\Vert \sqrt{\mathrm{Tr}\left(\boldsymbol{\Sigma}\left(n\right)\right)-\mathrm{Tr}\left(\mathbf{W}\right)}+\sqrt{\mathrm{Tr}\left(\mathbf{W}\right)}\right)^{2}.
\end{align}
Let $\frac{\mathbb{E}\left[\left.\left\Vert \mathbf{x}\left(n\right)\right\Vert ^{2}\right|I_{c}\left(n-1\right)\right]}{L^{2}\left(n\right)}=\varepsilon,$
we get $L\left(n\right)=\frac{1}{\sqrt{\varepsilon}}\left(1+\left\Vert \mathbf{A}-\mathbf{B}\boldsymbol{\Psi}\mathbf{A}\right\Vert \Theta\right)(\left\Vert \mathbf{B}\boldsymbol{\Psi}\mathbf{A}\right\Vert \sqrt{\mathrm{Tr}\left(\boldsymbol{\Sigma}\left(n\right)\right)-\mathrm{Tr}\left(\mathbf{W}\right)}+\sqrt{\mathrm{Tr}\left(\mathbf{W}\right)})$,
hence Lemma (\ref{lemma-limiter-range}) is proved.

\subsection{\label{sub:Proof-of-pdf-property-2-1-1}Proof of Lemma \ref{lemma- conditional drift}}

Based on the energy queue dynamics (\ref{eq:energy-dya}), it follows:

\begin{align}
 & \mathbb{E}\left\{ \left.\frac{1}{2}\left[\left(E\left(n+1\right)-\theta\right)^{2}-\left(E\left(n\right)-\theta\right)^{2}\right]\right|\boldsymbol{\Sigma}\left(n\right)\right\} \nonumber \\
 & \leq\mathbb{E}\left[\left.\left(\theta-E\left(n\right)\right)M^{2}\mathrm{Tr}\left(\mathbf{F}^{H}(n)\mathbf{F}(n)\right)\tau\right|\boldsymbol{\Sigma}\left(n\right)\right]\nonumber \\
 & +\mathbb{E}\left[\left(\theta+\alpha\left(n\right)\right)^{2}\right].\label{eq:delta En}
\end{align}
Applying Lemma 4 in \cite{kammoun2014linearexp}, it follows:
\begin{align}
 & \boldsymbol{\Sigma}\left(n\right)\big(\mathbf{\mathbf{\mathbf{\tilde{F}}}}^{a}\left(n\right)\big)^{H}\Big(\mathbf{\mathbf{\mathbf{\tilde{F}}}}^{a}\left(n\right)\boldsymbol{\Sigma}\left(n\right)\big(\mathbf{\mathbf{\mathbf{\tilde{F}}}}^{a}\left(n\right)\big)^{H}+\mathbf{I}\Big)^{-1}\mathbf{\mathbf{\mathbf{\tilde{F}}}}^{a}\left(n\right)\boldsymbol{\Sigma}\left(n\right)\nonumber \\
 & =\boldsymbol{\Sigma}\left(n\right)-\Big(\big(\mathbf{\mathbf{\mathbf{\tilde{F}}}}^{a}\left(n\right)\big)^{H}\mathbf{\mathbf{\mathbf{\tilde{F}}}}^{a}\left(n\right)+\boldsymbol{\Sigma}\left(n\right)^{-1}\Big)^{-1}\label{eq:sigma-inter1}
\end{align}

Substituting (\ref{eq:sigma-inter1}) into (\ref{dyna-sigma}), it
follows: 
\begin{align}
 & \mathbb{E}\left\{ \left.\frac{1}{2}\left[\mathrm{Tr}\left(\boldsymbol{\Sigma}\left(n+1\right)\right)-\mathrm{Tr}\left(\boldsymbol{\Sigma}\left(n\right)\right)\right]\right|\boldsymbol{\Sigma}\left(n\right)\right\} \nonumber \\
 & \leq\frac{1}{2}\mathrm{Tr}(\mathbf{W})+\mathbb{E}\bigg\{\frac{\left\Vert \mathbf{A}\mathbf{A}^{T}\right\Vert }{2}\bigg[\varepsilon\mathrm{Tr}\left(\boldsymbol{\Sigma}\left(n\right)\right)\nonumber \\
 & +\mathrm{Tr}\bigg(2\frac{M^{2}}{L^{2}(n)}\mathrm{Re}\left\{ \mathbf{F}^{H}\left(n\right)\mathbf{H}^{H}(n)\mathbf{H}(n)\mathbf{F}\left(n\right)\right\} \nonumber \\
 & +\boldsymbol{\Sigma}\left(n\right)^{-1}\bigg)^{-1}\bigg]-\frac{1}{2}\mathrm{Tr}\left(\boldsymbol{\Sigma}\left(n\right)\right)\bigg|\boldsymbol{\Sigma}\left(n\right)\bigg\}\label{eq:sigma-inter2}
\end{align}

Adding up (\ref{eq:delta En}) and (\ref{eq:sigma-inter2}) it follows
inequality (\ref{eq:lya-drift}), and hence lemma \ref{lemma- conditional drift}
is proved.

\vspace{-0.5cm}

\subsection{\label{sub:Proof-of-pdf-property-2-1-1-1}Proof of Theorem \ref{solution}}

Let $\mathbf{G}(n)=\mathbf{U}^{H}\mathbf{F}\left(n\right),$ denote
$\mathbf{G}_{r}(n)=\mathrm{Re}\left\{ \mathbf{G}(n)\right\} $ and
$\mathbf{G}_{i}(n)=\mathrm{Im}\left\{ \mathbf{G}(n)\right\} $. Let
$\boldsymbol{\Pi}_{0}=\left[\begin{array}{cc}
\boldsymbol{\Pi} & \mathbf{0}\\
\mathbf{0} & \boldsymbol{\Pi}
\end{array}\right]$, $\widetilde{\mathbf{G}}\left(n\right)=\left[\begin{array}{c}
\mathbf{G}_{r}(n)\\
\mathbf{G}_{i}(n)
\end{array}\right]$. Define $f\left(\widetilde{\mathbf{G}}\right)=\frac{\left\Vert \mathbf{A}\mathbf{A}^{T}\right\Vert }{2}\mathrm{Tr}\left(2\widetilde{\mathbf{G}}^{T}\boldsymbol{\Pi}_{0}\widetilde{\mathbf{G}}+\boldsymbol{\Sigma}^{-1}\right)^{-1}+M^{2}\tau\left(\theta-E\right)\mathrm{Tr}\left(\widetilde{\mathbf{G}}^{T}\widetilde{\mathbf{G}}\right).$
For any given $\widetilde{\mathbf{G}}$, there is a decomposition
that $\widetilde{\mathbf{G}}=\widetilde{\mathbf{G}}_{1}+\widetilde{\mathbf{G}}_{2}$,
where $\mathrm{col}\left(\widetilde{\mathbf{G}}_{1}\right)\subseteq\mathrm{span}\left(\mathrm{col}\left(\boldsymbol{\Pi}_{0}\right)\right)$
and $\mathrm{col}\left(\widetilde{\mathbf{G}}_{2}\right)\subseteq\mathrm{span}\left(\mathrm{col}\left(\boldsymbol{\Pi}_{0}\right)\right)^{\perp}$,
and we have $f\left(\widetilde{\mathbf{G}}_{1}\right)\leq f\left(\widetilde{\mathbf{G}}\right)$.
Therefore, $\mathbf{G}^{\ast}(n)$ must satisfy $\mathrm{col}\left(\widetilde{\mathbf{G}}^{\ast}\left(n\right)\right)\subseteq\mathrm{span}\left(\mathrm{col}\left(\boldsymbol{\Pi}_{0}\right)\right)$,
i.e., $\mathrm{col}\left(\mathbf{F}^{\ast}\left(n\right)\right)\subseteq\mathrm{span}\left(\mathrm{col}\left(\mathbf{H}\right)\right)$.
Let $\boldsymbol{\widetilde{\Delta}}_{1}=\left[\begin{array}{cc}
\boldsymbol{\Delta}_{1} & \mathbf{0}\\
\mathbf{0} & \boldsymbol{\Delta}_{1}
\end{array}\right]$ where $\boldsymbol{\Delta}_{1}=\left[\begin{array}{cc}
\widetilde{\boldsymbol{\Pi}} & \mathbf{0}\\
\mathbf{0} & \mathbf{0}
\end{array}\right]$ and $\widetilde{\boldsymbol{\Pi}}$ is the leading principal minor
of order $\mathrm{min}\left(N_{s},N_{t}\right)$ of $\boldsymbol{\Pi}$.
Let $\boldsymbol{\widetilde{\Delta}}_{2}=\left[\begin{array}{cc}
\boldsymbol{\Delta}_{2} & \mathbf{0}\\
\mathbf{0} & \boldsymbol{\Delta}_{2}
\end{array}\right]$ where $\boldsymbol{\Delta}_{2}=\left[\begin{array}{cc}
\widetilde{\boldsymbol{\Pi}}^{-1} & \mathbf{0}\\
\mathbf{0} & \mathbf{0}
\end{array}\right]$. Let $\widetilde{\mathbf{S}}=\left[\begin{array}{cc}
\hat{\mathbf{S}} & \mathbf{0}\\
\mathbf{0} & \hat{\mathbf{S}}
\end{array}\right]$, where $\hat{\mathbf{S}}=\left[\begin{array}{cc}
\mathbf{S} & \mathbf{0}\\
\mathbf{0} & \mathbf{I}_{\left(N_{t}-K\right)\times\left(N_{t}-K\right)}
\end{array}\right]$. Let $\mathbf{X}(n)=\frac{M}{L(\boldsymbol{\Sigma})}\widetilde{\mathbf{S}}\boldsymbol{\Delta}_{1}\widetilde{\mathbf{G}}\left(n\right)$.
Then Problem 1 is transformed into the following problem:
\begin{align*}
\mathcal{P}_{1}:\underset{\mathrm{\mathbf{X}}\left(n\right)}{\mathrm{min}} & \ \ L^{2}(\boldsymbol{\Sigma})\mathrm{Tr}\left(\widetilde{\mathbf{S}}\boldsymbol{\widetilde{\Delta}}_{2}^{2}\widetilde{\mathbf{S}}^{T}\mathbf{X}(n)\mathbf{X}^{T}(n)\right)\tau\left(\theta-E\right)\\
 & +\frac{\left\Vert \mathbf{A}\mathbf{A}^{T}\right\Vert }{2}\mathrm{Tr}\left(2\mathbf{X}^{T}(n)\mathbf{X}(n)+\boldsymbol{\Sigma}^{-1}\right)^{-1}\\
s.t. & \ \ L^{2}(\boldsymbol{\Sigma})\mathrm{Tr}\left(\widetilde{\mathbf{S}}\boldsymbol{\widetilde{\Delta}}_{2}^{2}\widetilde{\mathbf{S}}^{T}\mathbf{X}(n)\mathbf{X}^{T}(n)\right)\tau\leq E.
\end{align*}
The associated KKT conditions are given by
\begin{align}
 & L^{2}(\boldsymbol{\Sigma})\mathrm{Tr}\left(\widetilde{\mathbf{S}}\boldsymbol{\widetilde{\Delta}}_{2}^{2}\widetilde{\mathbf{S}}^{T}\mathbf{X}^{\ast}(n)\left(\mathbf{X}^{\ast}(n)\right)^{T}\right)\tau-E\leq0;\label{eq:KKT-1}\\
 & \beta\geq0;\\
 & \beta\left[L^{2}(\boldsymbol{\Sigma})\mathrm{Tr}\left(\widetilde{\mathbf{S}}\boldsymbol{\widetilde{\Delta}}_{2}^{2}\widetilde{\mathbf{S}}^{T}\mathbf{X}(n)\mathbf{X}^{T}(n)\right)\tau-E\right]=0;\\
 & \left\Vert \mathbf{A}\mathbf{A}^{T}\right\Vert \mathbf{X}(n)\left(2\mathbf{X}^{T}(n)\mathbf{X}(n)+\boldsymbol{\Sigma}^{-1}\right)^{-2}\nonumber \\
 & +\left(\theta-E+\beta\right)\tau L^{2}(\boldsymbol{\Sigma})\widetilde{\mathbf{S}}\boldsymbol{\widetilde{\Delta}}_{2}^{2}\widetilde{\mathbf{S}}^{T}\mathbf{X}(n)=\mathbf{0}.\label{eq:KKT-4}
\end{align}

Note that for non-convex problems, the KKT condition is necessary
and in order to find the global optimum, we need to test all the solutions
that satisfy the KKT conditions. Therefore, without loss of optimality,
we restrict the feasible solution set of the optimization problem
to be the solutions that satisfy the KKT conditions (\ref{eq:KKT-1})-
(\ref{eq:KKT-4}). By exploiting the structural properties of the
solutions that satisfy the of the KKT conditions, the objective function
of the optimization problem can be further simplified. Let $\widetilde{\mathbf{X}}(n)=\widetilde{\mathbf{S}}^{T}\mathbf{X}(n)\mathbf{S}$,
the KKT condition (\ref{eq:KKT-1})- (\ref{eq:KKT-4}) are equivalent
to the following conditions (\ref{eq:kkt-1})- (\ref{eq:kkt-4}).
\begin{align}
 & L^{2}(\boldsymbol{\Sigma})\mathrm{Tr}\left(\boldsymbol{\widetilde{\Delta}}_{2}^{2}\widetilde{\mathbf{X}}(n)\widetilde{\mathbf{X}}^{T}(n)\right)\tau-E\leq0;\label{eq:kkt-1}\\
 & \beta\geq0;\\
 & \beta\left[L^{2}(\boldsymbol{\Sigma})\mathrm{Tr}\left(\boldsymbol{\widetilde{\Delta}}_{2}^{2}\widetilde{\mathbf{X}}^{T}(n)\widetilde{\mathbf{X}}(n)\right)\tau-E\right]=0;\\
 & \left(2\widetilde{\mathbf{X}}^{T}(n)\widetilde{\mathbf{X}}(n)+\boldsymbol{\Lambda}^{-1}\right)^{-2}\widetilde{\mathbf{X}}^{T}(n)\nonumber \\
 & =\frac{\left(\theta-E+\beta\right)\tau L^{2}(\boldsymbol{\Sigma})}{\left\Vert \mathbf{A}\mathbf{A}^{T}\right\Vert }\widetilde{\mathbf{X}}^{T}(n)\boldsymbol{\widetilde{\Delta}}_{2}^{2}.\label{eq:kkt-4}
\end{align}
It follows that every column of $\widetilde{\mathbf{X}}^{T}(n)$ that
corresponds to a nonzero diagonal element of $\boldsymbol{\widetilde{\Delta}}_{2}^{2}$
is an eigenvector of $\left(2\widetilde{\mathbf{X}}^{T}(n)\widetilde{\mathbf{X}}(n)+\boldsymbol{\Lambda}^{-1}\right)^{-2}$.
Hence, it follows
\begin{align}
 & \left(2\widetilde{\mathbf{X}}^{T}(n)\widetilde{\mathbf{X}}(n)+\boldsymbol{\Lambda}^{-1}\right)\widetilde{\mathbf{X}}^{T}(n)\widetilde{\mathbf{X}}(n)\nonumber \\
 & =\sqrt{\frac{\left\Vert \mathbf{A}\mathbf{A}^{T}\right\Vert }{\left(\theta-E+\beta\right)\tau L^{2}(\boldsymbol{\Sigma})}}\widetilde{\mathbf{X}}^{T}(n))\boldsymbol{\widetilde{\Delta}}_{1}\widetilde{\mathbf{X}}(n).
\end{align}
Since $\widetilde{\mathbf{X}}^{T}(n))\boldsymbol{\widetilde{\Delta}}_{1}\widetilde{\mathbf{X}}(n)$
is symmetric matrices, it follows $\left(2\widetilde{\mathbf{X}}^{T}(n)\widetilde{\mathbf{X}}(n)+\boldsymbol{\Lambda}^{-1}\right)$
and $\widetilde{\mathbf{X}}^{T}(n)\widetilde{\mathbf{X}}(n)$ commute,
and $\widetilde{\mathbf{X}}^{T}(n)\widetilde{\mathbf{X}}(n)$ and
$\boldsymbol{\Lambda}^{-1}$ commute. Therefore, $\widetilde{\mathbf{X}}^{T}(n)\widetilde{\mathbf{X}}(n)$
and $\boldsymbol{\Lambda}^{-1}$ are simultaneous diagonalizable,
i.e., there exist an unitary matrix $\mathbf{P}$, such that $\widetilde{\mathbf{X}}^{T}(n)\widetilde{\mathbf{X}}(n)=\mathbf{P}\mathrm{diag}\left(\sigma_{1},\ldots,\sigma_{K}\right)\mathbf{P}^{T},$
and $\boldsymbol{\Lambda}^{-1}=\mathrm{diag}\left(\lambda_{1},\ldots,\lambda_{K}\right)$,
where $\sigma_{1},\ldots,\sigma_{K}$ and $\lambda_{1},\ldots,\lambda_{K}$
are the eigenvalues of $\widetilde{\mathbf{X}}^{T}(n)\widetilde{\mathbf{X}}(n)$
and $\boldsymbol{\Lambda}^{-1}$, respectively. Hence, it follows
the value of the objective function of Problem $\mathcal{P}_{1}$
is given by: 
\begin{align}
 & L^{2}(\boldsymbol{\Sigma})\mathrm{Tr}\left(\widetilde{\mathbf{S}}\boldsymbol{\widetilde{\Delta}}_{2}^{2}\widetilde{\mathbf{S}}^{T}\mathbf{X}(n)\mathbf{X}^{T}(n)\right)\tau\left(\theta-E\right)\nonumber \\
 & +\frac{\left\Vert \mathbf{A}\mathbf{A}^{T}\right\Vert }{2}\mathrm{Tr}\left(2\mathbf{X}^{T}(n)\mathbf{X}(n)+\boldsymbol{\Sigma}^{-1}\right)^{-1}\nonumber \\
= & \left\Vert \mathbf{A}\mathbf{A}^{T}\right\Vert \sum_{i=1}^{K}\left[\left(2\sigma_{i}+\lambda_{i}\right)^{-2}\sigma_{i}\right]+\frac{\left\Vert \mathbf{A}\mathbf{A}^{T}\right\Vert }{2}\sum_{i=1}^{K}\left(2\sigma_{i}+\lambda_{i}\right)^{-1}.\label{eq:cvx-1}
\end{align}
Note the optimal solution $\widetilde{\mathbf{X}}(n)$ must satisfy
the KKT condition (\ref{eq:kkt-1})-(\ref{eq:kkt-4}). However, (\ref{eq:cvx-1})
implies that for any $\widetilde{\mathbf{X}}(n)$ satisfies the KKT
condition (\ref{eq:kkt-1})-(\ref{eq:kkt-4}), the value of the objective
function of of Problem $\mathcal{P}_{1}$ only depends on the eigenvalues
of $\widetilde{\mathbf{X}}^{T}(n)\widetilde{\mathbf{X}}(n)$, i.e.,
$\sigma_{1},\ldots,\sigma_{K},$ and independent of the structure
of $\widetilde{\mathbf{X}}(n)$. Hence, let $\widetilde{\mathbf{X}}(n)=\left[\begin{array}{c}
\widehat{\mathbf{X}}(n)\\
\mathbf{0}
\end{array}\right]$, where $\widehat{\mathbf{X}}(n)\in\mathbb{R}^{K\times K}$ is a diagonal
matrix, and let $\mathbf{Y}\left(n\right)=\widehat{\mathbf{X}}^{T}(n)\widehat{\mathbf{X}}(n)$.
And Problem $\mathcal{P}_{1}$ is equivalent to the following Problem
$\mathcal{P}_{2}:$
\begin{align*}
\mathcal{P}_{2}:\underset{\mathbf{Y}\left(n\right)}{\mathrm{min}} & \ \ L^{2}(\boldsymbol{\Sigma})\mathrm{Tr}\left(\boldsymbol{\Pi}_{K}^{-2}\mathbf{Y}\left(n\right)\right)\tau\left(\theta-E\right)\\
 & +\frac{\left\Vert \mathbf{A}\mathbf{A}^{T}\right\Vert }{2}\mathrm{Tr}\left(2\mathbf{Y}\left(n\right)+\boldsymbol{\Lambda}^{-1}\right)^{-1}\\
s.t. & \ \ L^{2}(\boldsymbol{\Sigma})\mathrm{Tr}\left(\boldsymbol{\Pi}_{K}^{-2}\mathbf{Y}\left(n\right)\right)\tau\leq E.
\end{align*}
Note that $\mathrm{Tr}\left(\boldsymbol{\Pi}_{K}^{-2}\mathbf{Y}\left(n\right)\right)$
is convex in $\mathbf{Y}\left(n\right)$, $\mathrm{Tr}\left(2\mathbf{Y}\left(n\right)+\boldsymbol{\Lambda}^{-1}\right)^{-1}$
is convex in $\mathbf{Y}\left(n\right)$. Hence the objective function
$L(\boldsymbol{\Sigma})\mathrm{Tr}\left(\boldsymbol{\Pi}_{K}^{-2}\mathbf{Y}\left(n\right)\right)\tau\left(\theta-E\right)+\frac{\left\Vert \mathbf{A}\mathbf{A}^{T}\right\Vert }{2}\mathrm{Tr}\left(2\mathbf{Y}\left(n\right)+\boldsymbol{\Lambda}^{-1}\right)^{-1}$
of Problem $\mathcal{P}_{2}$ is convex in $\mathbf{Y}\left(n\right)$.
The constraint of Problem $\mathcal{P}_{2}$ is also convex in $\mathbf{Y}\left(n\right).$
Therefore, Problem $\mathcal{P}_{2}$ is convex. Therefore, we conclude
that Problem 1 is equivalent to Problem $\mathcal{P}_{2}$ which is
a convex optimization problem. The optimal solution $\mathbf{Y}^{\ast}\left(n\right)$
is given by 
\begin{align}
\mathbf{Y}^{\ast}\left(n\right) & =\frac{1}{2}\left[\frac{\boldsymbol{\Pi}_{K}}{L\left(\boldsymbol{\Sigma}\right)}\sqrt{\frac{\left\Vert \mathbf{A}\mathbf{A}^{T}\right\Vert }{\left(\left[\theta-E\right]^{+}+\beta\right)\tau}}-\boldsymbol{\Lambda}^{-1}\right]^{+},
\end{align}
where $\beta=0$ if $L^{2}(\boldsymbol{\Sigma})\mathrm{Tr}\left(\boldsymbol{\Pi}_{K}^{-2}\mathbf{Y}^{*}\left(n\right)\right)\tau<E$,
else $\beta$ is chosen such that $L^{2}(\boldsymbol{\Sigma})\mathrm{Tr}\left(\boldsymbol{\Pi}_{K}^{-2}\mathbf{Y}^{*}\left(n\right)\right)\tau=E$.
Since $\widehat{\mathbf{X}^{\ast}}(n)=\left(\mathbf{Y}^{\ast}\left(n\right)\right)^{\frac{1}{2}},$
$\widetilde{\mathbf{X}}(n)=\left[\begin{array}{c}
\widehat{\mathbf{X}}(n)\\
\mathbf{0}
\end{array}\right]$ and $\mathbf{X}(n)=\widetilde{\mathbf{S}}\widetilde{\mathbf{X}}(n)\mathbf{S}^{T}$,
it follows $\mathbf{X}^{\ast}(n)=\left[\begin{array}{c}
\mathbf{S}\widetilde{\boldsymbol{\Lambda}}\mathbf{S}^{T}\\
\mathbf{0}
\end{array}\right],$ where 
\begin{align}
 & \widetilde{\boldsymbol{\Lambda}}=\Big(\frac{1}{2}\Big[\frac{\boldsymbol{\Pi}_{K}}{L\left(\boldsymbol{\Sigma}\right)}\sqrt{\frac{\left\Vert \mathbf{A}\mathbf{A}^{T}\right\Vert }{\left(\left[\theta-E\right]^{+}+\beta\right)\tau}}-\boldsymbol{\Lambda}^{-1}\Big]^{+}\Big)^{\nicefrac{1}{2}}.
\end{align}
Therefore, $\mathbf{F}^{\ast}\left(n\right)=\frac{L(\boldsymbol{\Sigma})}{M}\mathbf{U}\left[\begin{array}{c}
\boldsymbol{\Pi}_{K}^{-1}\widetilde{\boldsymbol{\Lambda}}\mathbf{S}^{T}\\
\mathbf{0}
\end{array}\right]$, and if $L^{2}(\boldsymbol{\Sigma})\mathrm{Tr}\Big(\frac{1}{2}\boldsymbol{\Pi}_{K}^{-2}\Big[\frac{\boldsymbol{\Pi}_{K}}{L\left(\boldsymbol{\Sigma}\right)}\sqrt{\frac{\left\Vert \mathbf{A}\mathbf{A}^{T}\right\Vert }{\left(\theta-E\right)\tau}}\mathbf{I}-\boldsymbol{\Lambda}^{-1}\Big]^{+}\Big)\tau<E$
then $\beta=0$, else $\beta$ is chosen such that $M^{2}\mathrm{Tr}\left(\left(\mathbf{F}^{\ast}(n)\right)^{H}\mathbf{F}^{\ast}(n)\right)\tau=E$.
Therefore, Theorem is \ref{solution} proved.\vspace{-0.5cm}

\subsection{\label{sub:Proof-of-pdf-property-2-1}Proof of Theorem \ref{Thm suff condition}}

Intuitively, the negative Lyapunov drift is related to the stability
of the MIMO plant. Therefore, the sufficient condition for stability
should be the sufficient condition to guarantee negative Lyapunov
drift. Substituting the MIMO precoding solution in Theorem \ref{solution}
into (\ref{eq:sigma-inter2}), it follows:
\begin{align}
 & \mathrm{Tr}\left(2\frac{M^{2}}{L^{2}(n)}\mathrm{Re}\left\{ \mathbf{F}^{H}\left(n\right)\mathbf{H}^{H}(n)\mathbf{H}(n)\mathbf{F}\left(n\right)\right\} +\boldsymbol{\Sigma}\left(n\right)^{-1}\right)^{-1}\nonumber \\
= & \mathrm{Tr}\left(\left[\frac{\boldsymbol{\Pi}_{K}}{L\left(\boldsymbol{\Sigma}\right)}\sqrt{\frac{\left\Vert \mathbf{A}\mathbf{A}^{T}\right\Vert }{\left(\left[\theta-E\right]^{+}+\beta\right)\tau}}-\boldsymbol{\Lambda}^{-1}\right]^{+}+\boldsymbol{\Lambda}^{-1}\right)^{-1}\nonumber \\
\leq & \mathrm{Tr}\left(\left[\frac{\boldsymbol{\Pi}_{K}E}{\mathrm{Tr}\left(\boldsymbol{\Pi}_{K}^{-1}\right)\tau L^{2}\left(\boldsymbol{\Sigma}\right)\mathcal{M}\left(\mathbf{A}\right)}-\lambda_{min}\mathbf{I}\right]^{+}+\lambda_{min}\mathbf{I}\right)^{-1}\nonumber \\
 & +\frac{\theta^{2}}{\left\Vert \mathbf{A}\mathbf{A}^{T}\right\Vert }\nonumber \\
= & K\mathrm{min}\left\{ \tilde{\pi}^{-1}\frac{\tau L^{2}\left(\boldsymbol{\Sigma}\right)\mathcal{M}\left(\mathbf{A}\right)}{E},\lambda_{min}^{-1}\right\} +\frac{\theta^{2}}{\left\Vert \mathbf{A}\mathbf{A}^{T}\right\Vert },\label{eq:drift=00003Dinter1}
\end{align}
where $\tilde{\pi}=\frac{\pi}{\mathrm{Tr}\left(\boldsymbol{\Pi}_{K}^{-1}\right)}$
and $\pi$ is the unordered singular value of $\mathbf{H}$ and $\lambda_{min}$
is the minimum nonzero element of $\boldsymbol{\Lambda}^{-1}$. Furthermore,
\begin{align*}
 & \mathbb{E}\left[\left.K\mathrm{min}\left\{ \tilde{\pi}^{-1}\frac{\tau L^{2}\left(\boldsymbol{\Sigma}\right)\mathcal{M}\left(\mathbf{A}\right)}{E},\lambda_{min}^{-1}\right\} \right|\boldsymbol{\Sigma}\left(n\right)\right]\\
\leq & \mathbb{E}\left[\left.K\mathrm{min}\left\{ \tilde{\pi}^{-1}\frac{\tau L^{2}\left(\boldsymbol{\Sigma}\right)\mathcal{M}\left(\mathbf{A}\right)}{E},\mathrm{Tr}\left(\boldsymbol{\Sigma}\left(n\right)\right)\right\} \right|\boldsymbol{\Sigma}\left(n\right)\right],
\end{align*}
where the expectation is taken w.r.t. the randomness of $\tilde{\pi}$
and $E$. Note that $\tilde{\pi}$ is independent of $\boldsymbol{\Sigma}\left(n\right)$
and $\alpha\left(n\right)$, and there exists a constant $\xi$ such
that 
\begin{align}
 & \mathbb{E}_{\tilde{\pi},E}\left[\left.K\mathrm{min}\left\{ \tilde{\pi}^{-1}\frac{\tau L^{2}\left(\boldsymbol{\Sigma}\right)\mathcal{M}\left(\mathbf{A}\right)}{E},\mathrm{Tr}\left(\boldsymbol{\Sigma}\left(n\right)\right)\right\} \right|\boldsymbol{\Sigma}\left(n\right)\right]\nonumber \\
\leq & K\mathrm{Pr}\left(\tilde{\pi}<\xi\right)\mathrm{Tr}\left(\boldsymbol{\Sigma}\left(n\right)\right)\nonumber \\
 & +K\tau L^{2}\left(\boldsymbol{\Sigma}\right)\mathcal{M}\left(\mathbf{A}\right)\mathbb{E}\left[\left.\tilde{\pi}^{-1}\right|\tilde{\pi}\geq\xi\right]\mathbb{E}\left[\left.\frac{1}{E}\right|\boldsymbol{\Sigma}\left(n\right)\right]\label{eq:final-drift-1}
\end{align}

Furthermore, let $\delta=\sqrt{\frac{2}{\varepsilon}}\left(1+\left\Vert \mathbf{A}-\mathbf{B}\boldsymbol{\Psi}\mathbf{A}\right\Vert \Theta\right)\left\Vert \mathbf{B}\boldsymbol{\Psi}\right\Vert \left\Vert \mathbf{A}\right\Vert $,
it follows: 
\begin{align}
 & L^{2}\left(n\right)\leq\frac{1}{\varepsilon}\left(1+\left\Vert \mathbf{A}-\mathbf{B}\boldsymbol{\Psi}\mathbf{A}\right\Vert \Theta\right)^{2}\nonumber \\
 & \cdot(\left\Vert \mathbf{B}\boldsymbol{\Psi}\mathbf{A}\right\Vert \sqrt{\mathrm{Tr}\left(\boldsymbol{\Sigma}\left(n\right)\right)}+\sqrt{\mathrm{Tr}\left(\mathbf{W}\right)})^{2}\nonumber \\
 & \leq\frac{\delta^{2}}{\left\Vert \mathbf{B}\boldsymbol{\Psi}\right\Vert ^{2}\left\Vert \mathbf{A}\right\Vert ^{2}}\left(\left\Vert \mathbf{B}\boldsymbol{\Psi}\mathbf{A}\right\Vert ^{2}\mathrm{Tr}\left(\boldsymbol{\Sigma}\left(n\right)\right)+\mathrm{Tr}\left(\mathbf{W}\right)\right)\label{eq:final-L(n)}
\end{align}
 Note that if $\left[E\left(n-1\right)-\|\mathbf{F}\left(n-1\right)\mathbf{q}\left(n-1\right)\|^{2}\tau\right]^{+}+\alpha\left(n-1\right)<\theta$,
then $E>\alpha\left(n-1\right)$. Since $\alpha\left(n\right)$ is
i.i.d. energy arrival and independent of $\boldsymbol{\Sigma}\left(n\right)$,
it follows $\mathbb{E}\left[\left.\frac{1}{E}\right|\boldsymbol{\Sigma}\left(n\right)\right]<\mathbb{E}\left[\frac{1}{\alpha}\right]$.
If $\left[E\left(n-1\right)-\|\mathbf{F}\left(n-1\right)\mathbf{q}\left(n-1\right)\|^{2}\tau\right]^{+}+\alpha\left(n-1\right)\geq\theta$,
then $E=\theta$ and $\mathbb{E}\left[\left.\frac{1}{E}\right|\boldsymbol{\Sigma}\left(n\right)\right]=\frac{1}{\theta}$.
Therefore $\mathbb{E}\left[\left.\frac{1}{E}\right|\boldsymbol{\Sigma}\left(n\right)\right]\leq\mathrm{Pr}([E\left(n-1\right)-\|\mathbf{F}\left(n-1\right)\mathbf{q}\left(n-1\right)\|^{2}\tau]^{+}+\alpha\left(n-1\right)<\theta|\boldsymbol{\Sigma}\left(n\right))\mathbb{E}\left[\frac{1}{\alpha}\right]+\mathrm{Pr}([E\left(n-1\right)-\|\mathbf{F}\left(n-1\right)\mathbf{q}\left(n-1\right)\|^{2}\tau]^{+}+\alpha\left(n-1\right)\geq\theta|\boldsymbol{\Sigma}\left(n\right))\frac{1}{\theta}<\mathbb{E}\left[\frac{1}{\alpha}\right]+\frac{1}{\theta}$.
Substituting (\ref{eq:drift=00003Dinter1}), (\ref{eq:final-drift-1})
and (\ref{eq:final-L(n)}) into the drift of $\boldsymbol{\Sigma}\left(n\right)$
(\ref{eq:sigma-inter2}), it follows: 
\begin{align}
 & \mathbb{E}\left\{ \left.\left[\mathrm{Tr}\left(\boldsymbol{\Sigma}\left(n+1\right)\right)-\mathrm{Tr}\left(\boldsymbol{\Sigma}\left(n\right)\right)\right]\right|\boldsymbol{\Sigma}\left(n\right)\right\} \nonumber \\
 & \leq\bigg(1+K\tau\frac{\delta^{2}}{\left\Vert \mathbf{B}\boldsymbol{\Psi}\right\Vert ^{2}}\mathcal{M}\left(\mathbf{A}\mathbf{A}^{T}\right)\mathcal{M}\left(\mathbf{A}\right)\mathbb{E}\left[\left.\tilde{\pi}^{-1}\right|\tilde{\pi}\geq\xi\right]\nonumber \\
 & \cdot\left(\mathbb{E}\left[\frac{1}{\alpha}\right]+\frac{1}{\theta}\right)\bigg)\mathrm{Tr}\left(\mathbf{W}\right)+\theta^{2}+\mathrm{Tr}\left(\boldsymbol{\Sigma}\left(n\right)\right)\bigg\{\varepsilon\mathcal{M}\left(\mathbf{A}\mathbf{A}^{T}\right)\nonumber \\
 & +K\mathrm{Pr}\left(\tilde{\pi}<\xi\right)\mathcal{M}\left(\mathbf{A}\mathbf{A}^{T}\right)+\delta^{2}K\left(\mathbb{E}\left[\frac{1}{\alpha}\right]+\frac{1}{\theta}\right)\tau\nonumber \\
 & \cdot\mathcal{M}\left(\mathbf{A}\mathbf{A}^{T}\right)\mathcal{M}\left(\mathbf{A}\right)\mathbb{E}\left[\left.\tilde{\pi}^{-1}\right|\tilde{\pi}\geq\xi\right]-1\bigg\}.\label{eq:sigma-final}
\end{align}

Therefore, a sufficient condition for stability is given by $\varepsilon\mathcal{M}\left(\mathbf{A}\mathbf{A}^{T}\right)+K\mathrm{Pr}\left(\tilde{\pi}<\xi\right)\mathcal{M}\left(\mathbf{A}\mathbf{A}^{T}\right)+\delta^{2}K\left(\mathbb{E}\left[\frac{1}{\alpha}\right]+\frac{1}{\theta}\right)\tau\mathcal{M}\left(\mathbf{A}\mathbf{A}^{T}\right)\mathcal{M}\left(\mathbf{A}\right)\mathbb{E}\left[\left.\tilde{\pi}^{-1}\right|\tilde{\pi}\geq\xi\right]-1<0,$
i.e., 
\begin{align}
 & \mathbb{E}\left[\frac{1}{\alpha}\right]+\frac{1}{\theta}<\frac{1-\left(\varepsilon+K\mathrm{Pr}\left(\tilde{\pi}<\xi\right)\right)\mathcal{M}\left(\mathbf{A}\mathbf{A}^{T}\right)}{\delta^{2}K\tau\mathbb{E}\left[\left.\tilde{\pi}^{-1}\right|\tilde{\pi}\geq\xi\right]\mathcal{M}\left(\mathbf{A}\right)\mathcal{M}\left(\mathbf{A}\mathbf{A}^{T}\right)}.\label{eq:suff-condition}
\end{align}

Since there exists a constant $\xi$ such that (\ref{eq:final-drift-1})
holds, and for that typical $\xi$ there is a sufficient condition
in the form of (\ref{eq:suff-condition}) . Therefore, $\xi$ is chosen
as $\xi^{\ast}=\underset{\xi}{\mathrm{arg}\ \mathrm{max}}\ \frac{1-\left(\varepsilon+K\mathrm{Pr}\left(\tilde{\pi}<\xi\right)\right)\mathcal{M}\left(\mathbf{A}\mathbf{A}^{T}\right)}{K\delta^{2}\tau\mathbb{E}\left[\left.\tilde{\pi}^{-1}\right|\tilde{\pi}\geq\xi\right]\mathcal{M}\left(\mathbf{A}\right)\mathcal{M}\left(\mathbf{A}\mathbf{A}^{T}\right)}$,
and hence Theorem \ref{Thm suff condition} is proved.\vspace{-0.1cm}

\subsection{\label{sub:Proof-of-pdf-property-1}Proof of Theorem \ref{Thm MSE bound}}

If the sufficient condition of stability (\ref{eq:suff condition})
is satisfied, then denote a positive constant $\eta$ as $\eta=1-(\varepsilon+K\mathrm{Pr}\left(\tilde{\pi}<\xi^{\ast}\right))\mathcal{M}\left(\mathbf{A}\mathbf{A}^{T}\right)-\left(\mathbb{E}\left[\frac{1}{\alpha}\right]+\frac{1}{\theta}\right)KT\delta^{2}\tau\mathbb{E}\left[\left.\tilde{\pi}^{-1}\right|\tilde{\pi}\geq\xi^{\ast}\right]\mathcal{M}\left(\mathbf{A}\right)\mathcal{M}\left(\mathbf{A}\mathbf{A}^{T}\right).$
Substituting $\eta$ into equation (\ref{eq:sigma-final}), it follows:
\begin{align}
 & \mathbb{E}\left\{ \left.\left[\mathrm{Tr}\left(\boldsymbol{\Sigma}\left(n+1\right)\right)-\mathrm{Tr}\left(\boldsymbol{\Sigma}\left(n\right)\right)\right]\right|\boldsymbol{\Sigma}\left(n\right)\right\} \nonumber \\
\leq & (1+K\tau\frac{\delta^{2}}{\left\Vert \mathbf{B}\boldsymbol{\Psi}\right\Vert ^{2}}\mathcal{M}\left(\mathbf{A}\right)\mathcal{M}\left(\mathbf{A}\mathbf{A}^{T}\right)\mathbb{E}\left[\left.\tilde{\pi}^{-1}\right|\tilde{\pi}\geq\xi\right]\nonumber \\
 & \cdot\left(\mathbb{E}\left[\frac{1}{\alpha}\right]+\frac{1}{\theta}\right))\mathrm{Tr}\left(\mathbf{W}\right)+\theta^{2}-\eta\mathrm{Tr}\left(\boldsymbol{\Sigma}\left(n\right)\right).\label{eq:take expectation}
\end{align}
Taking expectation w.r.t. $\boldsymbol{\Sigma}\left(n\right)$ at
both side of (\ref{eq:take expectation}) and using the law of iterated
expectations yields:
\begin{align}
 & \mathbb{E}\left[\mathrm{Tr}\left(\boldsymbol{\Sigma}\left(n+1\right)\right)-\mathrm{Tr}\left(\boldsymbol{\Sigma}\left(n\right)\right)\right]\nonumber \\
\leq & (1+K\tau\frac{\delta^{2}}{\left\Vert \mathbf{B}\boldsymbol{\Psi}\right\Vert ^{2}}\mathcal{M}\left(\mathbf{A}\right)\mathcal{M}\left(\mathbf{A}\mathbf{A}^{T}\right)\mathbb{E}\left[\left.\tilde{\pi}^{-1}\right|\tilde{\pi}\geq\xi\right]\nonumber \\
 & \cdot\left(\mathbb{E}\left[\frac{1}{\alpha}\right]+\frac{1}{\theta}\right))\mathrm{Tr}\left(\mathbf{W}\right)+\theta^{2}-\eta\mathbb{E}\left[\mathrm{Tr}\left(\boldsymbol{\Sigma}\left(n\right)\right)\right].
\end{align}
Summing over time sots $n\in\left\{ 0,\ldots,N-1\right\} $ and dividing
by $N$ yields 
\begin{align}
 & \frac{\mathbb{E}\left[\mathrm{Tr}\left(\boldsymbol{\Sigma}\left(N\right)\right)-\mathrm{Tr}\left(\boldsymbol{\Sigma}\left(0\right)\right)\right]}{N}\nonumber \\
 & \leq(1+K\tau\frac{\delta^{2}}{\left\Vert \mathbf{B}\boldsymbol{\Psi}\right\Vert ^{2}}\mathcal{M}\left(\mathbf{A}\right)\mathcal{M}\left(\mathbf{A}\mathbf{A}^{T}\right)\mathbb{E}\left[\left.\tilde{\pi}^{-1}\right|\tilde{\pi}\geq\xi\right]\nonumber \\
 & \cdot\left(\mathbb{E}\left[\frac{1}{\alpha}\right]+\frac{1}{\theta}\right))\mathrm{Tr}\left(\mathbf{W}\right)+\theta^{2}-\frac{\eta}{N}\sum_{n=0}^{N-1}\mathbb{E}\left[\mathrm{Tr}\left(\boldsymbol{\Sigma}\left(n\right)\right)\right].\label{eq:summation}
\end{align}
By non-negativity of $\mathrm{Tr}\left(\boldsymbol{\Sigma}\left(n\right)\right)$,
it follows:
\begin{align}
 & \frac{1}{N}\sum_{n=0}^{N-1}\mathbb{E}\left[\mathrm{Tr}\left(\boldsymbol{\Sigma}\left(n\right)\right)\right]\nonumber \\
 & \leq\frac{1}{\eta}(1+K\tau\frac{\delta^{2}}{\left\Vert \mathbf{B}\boldsymbol{\Psi}\right\Vert ^{2}}\mathcal{M}\left(\mathbf{A}\right)\mathcal{M}\left(\mathbf{A}\mathbf{A}^{T}\right)\mathbb{E}\left[\left.\tilde{\pi}^{-1}\right|\tilde{\pi}\geq\xi\right]\nonumber \\
 & \cdot\left(\mathbb{E}\left[\frac{1}{\alpha}\right]+\frac{1}{\theta}\right))\mathrm{Tr}\left(\mathbf{W}\right)+\frac{\theta^{2}}{\eta}.\label{eq:final bound}
\end{align}
Taking limits of the inequality (\ref{eq:final bound}) as $N\rightarrow\infty$
yields 
\begin{align}
 & \limsup_{N\rightarrow\infty}\frac{1}{N}\sum_{n=0}^{N-1}\mathbb{E}\left[\mathrm{Tr}\left(\boldsymbol{\Sigma}\left(n\right)\right)\right]\nonumber \\
 & \leq\frac{1}{\eta}(1+K\tau\frac{\delta^{2}}{\left\Vert \mathbf{B}\boldsymbol{\Psi}\right\Vert ^{2}}\mathcal{M}\left(\mathbf{A}\right)\mathcal{M}\left(\mathbf{A}\mathbf{A}^{T}\right)\mathbb{E}\left[\left.\tilde{\pi}^{-1}\right|\tilde{\pi}\geq\xi\right]\nonumber \\
 & \cdot\left(\mathbb{E}\left[\frac{1}{\alpha}\right]+\frac{1}{\theta}\right))\mathrm{Tr}\left(\mathbf{W}\right)+\frac{\theta^{2}}{\eta}.
\end{align}

Note that from Lemma \ref{lemma-mse-bound}, we have $\underset{N\rightarrow\infty}{\mathrm{limsup}}\frac{1}{N}\sum_{n=0}^{N-1}\mathbb{E}[\left\Vert \mathbf{x}\left(n\right)-\hat{\mathbf{x}}\left(n\right)\right\Vert ^{2}]\leq\underset{N\rightarrow\infty}{\mathrm{limsup}}\frac{1}{N}\sum_{n=0}^{N-1}\mathbb{E}[\mathrm{Tr}\left(\boldsymbol{\Sigma}\left(n\right)\right)]$,
hence Theorem \ref{Thm MSE bound} is proved.


\begin{thebibliography}{10}
\bibitem{ftnote1} A. Bemporad, M. Heemels, and M. Johansson, \textit{Networked
Control Systems}. Springer, 2010, vol. 406.

\bibitem{lya3} L. Tassiulas and A. Ephremides, \textquotedblleft Stability
properties of constrained queueing systems and scheduling policies
for maximum throughput in multihop radio networks,\textquotedblright{}
\textit{IEEE Trans. Autom. Control}, vol. 37, no. 12, pp. 1936\textendash 1948,
1992.

\bibitem{lya7} Y. Cui, V. K. Lau, R. Wang, H. Huang, and S. Zhang,
\textquotedblleft A survey on delay-aware resource control for wireless
systemsa\k{ }\l large deviation theory, stochastic lyapunov drift,
and distributed stochastic learning,\textquotedblright{} \textit{IEEE
Trans. Inf. Theory}, vol. 58, no. 3, pp. 1677\textendash 1701, 2012.

\bibitem{lya8} P. Kumar and S. P. Meyn, \textquotedblleft Duality
and linear programs for stability and performance analysis of queuing
networks and scheduling policies,\textquotedblright{} \textit{IEEE
Trans. Autom. Control}, vol. 41, no. 1, pp. 4\textendash 17, 1996.

\bibitem{mmse-precoder} E. Biglieri, R. Calderbank, A. Constantinides,
A. Goldsmith, A. Paulraj, and H. V. Poor, \textit{MIMO Wireless Communications}.
Cambridge university press, 2007.

\bibitem{Sampath2001} H. Sampath, P. Stoica, and A. Paulraj, \textquotedblleft Generalized
linear precoder and decoder design for mimo channels using the weighted
mmse criterion,\textquotedblright{} \textit{IEEE Trans. Commun.},
vol. 49, no. 12, pp. 2198\textendash 2206, 2001.

\bibitem{wiesel2006linear} A. Wiesel, Y. C. Eldar, and S. Shamai,
\textquotedblleft Linear precoding via conic optimization for fixed
MIMO receivers,\textquotedblright{} \textit{IEEE Trans. Signal Process.},
vol. 54, no. 1, pp. 161\textendash 176, 2006.

\bibitem{li2013optimal} Y. Li, D. E. Quevedo, V. Lau, and L. Shi,
\textquotedblleft Optimal periodic transmission power schedules for
remote estimation of arma processes,\textquotedblright{} \textit{IEEE
Trans. Signal Process.}, vol. 61, no. 24, pp. 6164\textendash 6174,
2013.

\bibitem{bertsekas1995dynamic} D. P. Bertsekas, \textit{Dynamic Programming
and Optimal Control}. Athena Scientific Belmont, MA, 1995, vol. 1,
no. 2.

\bibitem{event-driven1} Y. Xu and J. P. Hespanha, \textquotedblleft Estimation
under uncontrolled and controlled communications in networked control
systems,\textquotedblright{} in \textit{Decision and Control, 2005
and 2005 European Control Conference. CDC-ECC\textquoteright 05. 44th
IEEE Conference on. IEEE}, 2005, pp. 842\textendash 847.

\bibitem{event-driven2} A. Molin and S. Hirche, \textquotedblleft An
iterative algorithm for optimal eventtriggered estimation,\textquotedblright{}
in \textit{IFAC Conference on Analysis and Design of Hybrid Systems},
2012, pp. 64\textendash 69.

\bibitem{optimal-kf-power-allo} M. Nourian, A. S. Leong, and S. Dey,
\textquotedblleft Optimal energy allocation for kalman filtering over
packet dropping links with imperfect acknowledgments and energy harvesting
constraints,\textquotedblright{} \textit{IEEE Trans. Autom. Control},
vol. 59, no. 8, pp. 2128\textendash 2143, 2014.

\bibitem{optimal-strategy} A. Nayyar, T. Basar, D. Teneketzis, and
V. V. Veeravalli, \textquotedblleft Optimal strategies for communication
and remote estimation with an energy harvesting sensor,\textquotedblright{}
IEEE Trans. Autom. Control, vol. 58, no. 9, pp. 2246\textendash 2260,
2013.

\bibitem{sastry2013nonlinear} S. Sastry, \textit{Nonlinear Systems:
Analysis, Stability, and Control}. Springer Science \& Business Media,
2013, vol. 10.

\bibitem{neely2010stochasticlyaneely} M. J. Neely, \textquotedblleft Stochastic
network optimization with application to communication and queueing
systems,\textquotedblright{}\textit{ Synthesis Lectures on Communication
Networks}, vol. 3, no. 1, pp. 1\textendash 211, 2010.

\bibitem{aastrom2013adaptiveadptivectrl} K. J. Aström and B. Wittenmark,
\textit{Adaptive Control}. Courier Corporation, 2013.

\bibitem{CE1} S. Tatikonda, A. Sahai, and S. Mitter, \textquotedblleft Stochastic
linear control over a communication channel,\textquotedblright{}\textit{
IEEE Trans. Autom. Control}, vol. 49, no. 9, pp. 1549\textendash 1561,
2004.

\bibitem{CE2} J. P. Hespanha and A. S. Morse, \textquotedblleft Certainty
equivalence implies detectability,\textquotedblright{} \textit{Systems
\& Control Letters}, vol. 36, no. 1, pp. 1\textendash 13, 1999.

\bibitem{neely2005dynamiclyaopt} M. J. Neely, E. Modiano, and C.
E. Rohrs, \textquotedblleft Dynamic power allocation and routing for
time-varying wireless networks,\textquotedblright{} \textit{IEEE J.
Sel. Areas Commun.}, vol. 23, no. 1, pp. 89\textendash 103, 2005.

\bibitem{neely2006energy} M. J. Neely, \textquotedblleft Energy optimal
control for time-varying wireless networks,\textquotedblright{}\textit{
IEEE Trans. Inf. Theory}, vol. 52, no. 7, pp. 2915\textendash 2934,
2006.

\bibitem{epi-lemma1} T. M. Cover and J. A. Thomas, \textit{Elements
of Information Theory}. John Wiley \& Sons, 2012.

\bibitem{EH-precoder} S. Ulukus, A. Yener, E. Erkip, O. Simeone,
M. Zorzi, P. Grover, and K. Huang, \textquotedblleft Energy harvesting
wireless communications: A review of recent advances,\textquotedblright{}
\textit{IEEE J. Sel. Areas Commun.}, vol. 33, no. 3, pp. 360\textendash{}
381, 2015.

\bibitem{stein1989bode} G. Stein, \textquotedblleft Bode lecture:
respect the unstable,\textquotedblright{} in \textit{Proc. of Conference
on Decision and Control}, 1989.

\bibitem{qiu2010quantify} L. Qiu, \textquotedblleft Quantify the
unstable,\textquotedblright{} in \textit{Proc. of the 19th International
Symposium on Mathematical Theory of Networks and Systems MTNS}, 2010,
pp. 981\textendash 986.

\bibitem{stab-compare-1} G. N. Nair and R. J. Evans, \textquotedblleft Stabilizability
of stochastic linear systems with finite feedback data rates,\textquotedblright{}
\textit{SIAM Journal on Control and Optimization}, vol. 43, no. 2,
pp. 413\textendash 436, 2004.

\bibitem{stabi-compare-2} S. Tatikonda and S. Mitter, \textquotedblleft Control
over noisy channels,\textquotedblright{} \textit{IEEE Trans. Autom.
Control}, vol. 49, no. 7, pp. 1196\textendash 1201, 2004.

\bibitem{shi2007downlink} S. Shi, M. Schubert, and H. Boche, \textquotedblleft Downlink
MMSE transceiver optimization for multiuser mimo systems: Duality
and sum-mse minimization,\textquotedblright{} \textit{IEEE Trans.
Signal Process.}, vol. 55, no. 11, pp. 5436\textendash 5446, 2007.

\bibitem{sinopoli2004kalmanintermit} B. Sinopoli, L. Schenato, M.
Franceschetti, K. Poolla, M. Jordan, S. S. Sastry et al., \textquotedblleft Kalman
filtering with intermittent observations,\textquotedblright{} \textit{IEEE
Trans. Autom. Control}, vol. 49, no. 9, pp. 1453\textendash 1464,
2004.

\bibitem{augmented} S. L. Goh and D. P. Mandic, \textquotedblleft An
augmented extended kalman filter algorithm for complex-valued recurrent
neural networks,\textquotedblright{} \textit{Neural Computation},
vol. 19, no. 4, pp. 1039\textendash 1055, 2007.

\bibitem{O.Rioul} O. Rioul, \textquotedblleft Information theoretic
proofs of entropy power inequalities,\textquotedblright{} \textit{IEEE
Trans. Inf. Theory}, vol. 57, no. 1, pp. 33\textendash 55, 2011.

\bibitem{limiterproof} R. W. Brockett and D. Liberzon, \textquotedblleft Quantized
feedback stabilization of linear systems,\textquotedblright{} \textit{IEEE
Trans. Autom. Control}, vol. 45, no. 7, pp. 1279\textendash{} 1289,
2000.

\bibitem{kammoun2014linearexp} A. Kammoun, A. Muller, E. Bjornson,
and M. Debbah, \textquotedblleft Linear precoding based on polynomial
expansion: Large-scale multi-cell mimo systems,\textquotedblright{}
\textit{IEEE J. Sel. Topics Signal Process.}, vol. 8, no. 5, pp. 861\textendash{}
875, 2014.\end{thebibliography}
\end{document}